\documentclass[english,a4paper,twoside,openright,12pt, preprint]{aastex}
\usepackage[english]{babel}
\usepackage{natbib}
\usepackage[nomarkers,notablist,nofiglist]{endfloat}
\usepackage{longtable}
\usepackage{lscape}
\usepackage{multirow}
\usepackage{graphicx}
\usepackage{hyperref}
\usepackage{color}
\usepackage[figuresright]{rotating}
\usepackage{changepage}
\usepackage{booktabs}

\newcommand{\salttwo}{{\sc{SALT2}}}
\newcommand{\mlcs}{{\sc{MLCS2k2}}}
\newcommand{\mlcsshort}{{\sc{MLCS}}}
\newcommand{\lc}{light-curve}
\newcommand{\lccap}{Light-curve}

\shorttitle{SDSS-II SN properties vs GCD}
\shortauthors{Galbany et al.}

\begin{document}

\title{Type Ia Supernova Properties as a Function of the Distance to the Host Galaxy in the SDSS-II SN Survey}

\author{Llu\'is Galbany\altaffilmark{1,a}, 
 Ramon Miquel\altaffilmark{1,2}, 
 Linda \"{O}stman\altaffilmark{1}, 
 Peter J. Brown\altaffilmark{3}, 
 David Cinabro\altaffilmark{4}, 
 Chris B. D'Andrea\altaffilmark{5},
 Joshua Frieman\altaffilmark{6,7,8},
 Saurabh W. Jha\altaffilmark{9}
 John Marriner\altaffilmark{8}, 
 Robert C. Nichol\altaffilmark{5}, 
 Jakob Nordin\altaffilmark{10,11}, 
 Matthew D. Olmstead\altaffilmark{3}, 
 Masao Sako\altaffilmark{12},
 Donald P. Schneider\altaffilmark{13,14},
 Mathew Smith\altaffilmark{15}, 
 Jesper Sollerman\altaffilmark{16},
 Kaike Pan\altaffilmark{17}, 
 Stephanie Snedden\altaffilmark{17}, 
 Dmitry Bizyaev\altaffilmark{17}, 
 Howard Brewington\altaffilmark{17}, 
 Elena Malanushenko\altaffilmark{17}, 
 Viktor Malanushenko\altaffilmark{17}, 
 Dan Oravetz\altaffilmark{17}, 
 Audrey Simmons\altaffilmark{17}, 
 and Alaina Shelden\altaffilmark{17}}

\altaffiltext{1}{Institut de F\'{i}sica d'Altes Energies, Universitat Aut\`{o}noma de Barcelona, 
E-08193 Bellaterra (Barcelona), Spain}
\altaffiltext{2}{Instituci\'o Catalana de Recerca i Estudis Avan\c{c}ats, E-08010 Barcelona, Spain}
\altaffiltext{3}{Department of Physics and Astronomy, University of Utah, Salt Lake City, UT 84112, USA.}
\altaffiltext{4}{Wayne State University, Department of Physics and Astronomy, Detroit, MI, 48201, USA.}
\altaffiltext{5}{Institute of Cosmology and Gravitation, University of Portsmouth, Dennis Sciama Building, Burnaby Road, Portsmouth, PO1 3FX, UK.}
\altaffiltext{6}{Kavli Institute for Cosmological Physics, The University of Chicago, 5640 South Ellise Avenue, Chicago,
IL 60637, USA}
\altaffiltext{7}{Department of Astronomy and Astrophysics, The University of Chicago, 5640 South Ellise Avenue,
Chicago, IL 60637, USA}
\altaffiltext{8}{Center for Astrophysics, Fermi National Accelerator Laboratory, P.O. Box 500, Batavia, IL 60510, USA.}
\altaffiltext{9}{Department of Physics and Astronomy, Rutgers the State University of New Jersey, 136 Frelinghuysen Road, Piscataway, NJ 08854, USA.}
\altaffiltext{10}{E.O. Lawrence Berkeley National Lab, 1 Cyclotron Rd., Berkeley, CA 94720.}
\altaffiltext{11}{Space Sciences Laboratory, University of California Berkeley, Berkeley, CA 94720.}
\altaffiltext{12}{University of Pennsylvania, 209 South 33rd Street, Philadelphia, PA 19104}
\altaffiltext{13}{Department of Astronomy and Astrophysics, The Pennsylvania State University, University Park, PA 16802, USA.}
\altaffiltext{14}{Institute for Gravitation and the Cosmos, The Pennsylvania State University, University Park, PA 16802, USA.}
\altaffiltext{15}{Department of Physics, University of Western Cape, Bellville 7535, Cape Town, South Africa.}
\altaffiltext{16}{The Oskar Klein Centre, Department of Astronomy, AlbaNova, SE-106 91 Stockholm, Sweden.}
\altaffiltext{17}{Apache Point Observatory, P.O. Box 59, Sunspot, NM 88349, USA.}
\altaffiltext{a}{Now at CENTRA - Centro Multidisciplinar de Astrof\'isica, Instituto Superior T\'ecnico, Av. Rovisco Pais 1, 1049-001 Lisbon, Portugal. Email: lluis.galbany@ist.utl.pt}
\begin{abstract}
We use type-Ia supernovae (SNe Ia) discovered by the SDSS-II SN Survey to search for dependencies between SN Ia properties and the projected distance to the host galaxy center, using the distance as a proxy for local galaxy properties (local star-formation rate, local metallicity, etc.). 
The sample consists of almost 200 spectroscopically or photometrically confirmed SNe Ia at redshifts below 0.25. The sample is split into two groups depending on the morphology of the host galaxy.
We fit {\lc}s using both {\mlcs} and {\salttwo}, and determine color ($A_V$, $c$) and {\lc} shape ($\Delta$, $x_1$) parameters for each SN Ia, as well as its residual in the Hubble diagram. We then correlate
these parameters with both the physical and the normalized distances to the center of the host galaxy and look for trends in the mean values and scatters of these parameters with increasing distance. 
The most significant (at the 4~$\sigma$ level) finding is that the average fitted $A_V$ from {\mlcs} and $c$ from {\salttwo} decrease with the projected distance for SNe Ia in spiral galaxies. 
We also find indications that SNe in elliptical galaxies tend to have narrower light-curves if they explode at larger distances, although this may be due to selection effects in our sample. We do not find strong correlations between the residuals of the distance moduli with respect to the Hubble flow and the galactocentric distances, which indicates a limited correlation between SN magnitudes after standardization and local host metallicity.
\end{abstract}

\section{Introduction}
In 1998, the study of the redshift-luminosity relation (Hubble diagram) for nearby and distant Type Ia supernovae (SNe Ia) \citep{Riess:1998p100,Perlmutter:1999p180} provided the ``smoking gun'' for the accelerated expansion of the universe.
Since then, several surveys have added substantial statistics to the Hubble diagram \citep[e.g.][]{Astier:2006p513,WoodVasey:2007p581,Hicken:2009p217,Kessler:2009p310,Conley:2011p128} and extended it to higher redshifts \citep[e.g.][]{Knop:2003p103,Tonry:2003p5134,Riess:2004p260,Barris:2004p8616,Amanullah:2010p545,Suzuki:2011p2199}, thus strengthening the evidence for the accelerating universe.

SNe Ia can serve as cosmological probes because of their ability to function as reliable and accurate distance indicators on cosmological scales. This ability rests on the empirical correlation between the SN peak brightness and {\lc} width \citep{Phillips:1993p292}. Several empirical techniques \citep{Riess:1996p490, Phillips:1999p499, Barris:2004p7, Guy:2005p86, Prieto:2006p149, Jha:2007p262, Guy:2007p528} have been developed to exploit this correlation and turn SNe Ia into standard candles, with a dispersion on their corrected peak magnitude of 0.10--0.15~mag, corresponding to a precision of $\sim$5--7\% in distance.

As both the quantity and quality of supernova observations have increased, limitations of the homogeneity of SNe Ia have become apparent \citep{Riess:1996p490, Sullivan:2006p504}. If these inhomogeneities are not accounted for by the {\lc} width and color corrections or by other means, these variations may introduce systematic errors in the determination of cosmological parameters from supernova surveys. One plausible source of inhomogeneity is a dependence of supernova properties on host galaxy features. Since the average properties of host galaxies evolve with redshift, any such dependence will impact the cosmological parameter determination.
There have been many recent studies illustrating the dependence of SN properties on global characteristics of their hosts \citep{Sullivan:2006p504, Gallagher:2008p27, Howell:2009p544, Hicken:2009p326, Kelly:2010p599, Sullivan:2010p113}, also by the SDSS-II SN collaboration \citep{Lampeitl:2010p317,Smith:2011p8634,2011ApJ...743..172D,Gupta:2011p4678,Nordin:2011p50,Konishi:2011p353}. Much has been learned from these studies. For instance, it has by now been established \citep{Hamuy:1996p601, Gallagher:2005p1222, Sullivan:2006p504,Lampeitl:2010p317} that SNe Ia in passive galaxies are, on average, dimmer than those in star-forming galaxies. These SNe also have narrower {\lc}s,
and, after applying the light-curve standardization procedure, turn out to have slightly larger corrected peak brightnesses~\citep{Lampeitl:2010p317}

Following earlier work by \cite{2000ApJ...542..588I}, \citet{Jha:2006p80} and \citet{Hicken:2009p326}, we present here a study of the dependency of SN Ia properties with local characteristics of their host galaxies, using the location of the supernova inside the galaxy as a proxy for physically relevant parameters, such as local metallicity or local star-formation rate. We use the three-year Sloan Digital Sky Survey-II (SDSS-II) Supernova Survey sample \citep{Frieman:2008p236}, as well as the Fall 2004 test campaign sample, restricting the redshifts to $z<0.25$ in order to minimize observational biases. We examine the supernova {\lc} parameters related to color and decline rate, as well as the Hubble-diagram residuals, as a function of the projected distance between the supernova and the center of its host galaxy. We use the output parameters from two {\lc} fitters, {\mlcs} \citep{Jha:2007p262} and {\salttwo} \citep{Guy:2007p528}. For {\mlcs} we obtain $A_V$ as a measure of the color and $\Delta$ for the {\lc} width / decline rate. The corresponding parameter for {\salttwo} for color is $c$ and for {\lc} width, $x_1$. 

The outline of the paper is as follows. In Section 2 we describe the supernova sample and the host galaxy information used in the analysis. Section 3 covers the selection of SNe Ia, the procedure used for separating the host galaxies according to their morphology, and the description of the {\lc} parameters studied. In Section 4 we introduce  the method used to extract correlations between {\lc} parameters and distance to the host galaxy, and present the results of the analysis. Finally, in Section 5 we discuss these results, and offer some conclusions.
%
%
\section{Data Sample}
\subsection{SDSS-II Supernova Sample}
The Sloan Digital Sky Survey-II Supernova Survey \citep{Frieman:2008p236} has identified and measured {\lc}s for intermediate redshift ($0.01<z<0.45$) SNe during the three Fall seasons of operation from 2005 to 2007, using the dedicated SDSS 2.5m telescope at Apache Point Observatory \citep{Gunn:1998p1708,Gunn:2006p159}. A handful of supernovae were also obtained in the Fall 2004 test campaign.
The SNe are all located in Stripe 82, a 300~deg$^2$ region along the Celestial Equator in the Southern Galactic hemisphere \citep{Stoughton:2002p547}. The target selection is presented in \citet{Sako:2008p210}, the first year photometry in \citet{Holtzman:2008p373}, the first year spectroscopy in \citet{Zheng:2008p275}, and the second and third year NTT/NOT spectroscopy in \citet{Ostman:2011p331}. 
The SDSS-II SN survey has discovered and confirmed spectroscopically 559 SNe Ia, of which 514 were confirmed by the SDSS-II SN collaboration, 36 are likely SNe Ia, and 9 were confirmed by other groups. We will refer to these SNe as the ``Spec-Ia'' sample. Besides the spectroscopically confirmed SNe, the SDSS-II SN sample has 759 SNe photometrically classified as Type Ia from their {\lc}s, with spectroscopic redshifts of the host galaxy either measured previously by the SDSS Legacy Survey \citep{York:2000p440} or recently by the SDSS-III Baryon Oscillation Spectroscopic Survey (BOSS, see \citealp{2011AJ....142...72E} for an overview, and \citealp{Olmstead:2012} for the BOSS redshifts). 
The classification is based on the algorithm presented in~\citealp{Sako:2011p605}, which compares the SNe light-curves against a grid of SNe Ia light-curve models and core-collapse SNe light-curve templates, choosing the best-matching SN type using the host galaxy spectroscopic redshift as a prior.
We designate this SN sample as the ``Photo-Ia'' sample. The expected contamination of non-Ia SNe in the Photo-Ia sample is $\sim$6\% \citep{Sako:2011p605}. The number of SNe in the Photo-Ia sample has been significantly increased with the BOSS contribution. The entire SDSS-II SN sample, combining the Spec-Ia and Photo-Ia samples, consists of 1318 SNe Ia. Note that all these SNe have spectroscopically determined redshifts, either from the host galaxy or from the supernova spectrum.

Several host-galaxy analyses have been performed using the SDSS-II SN sample. \citet{Nordin:2011p446,Nordin:2011p50} and \citet{Konishi:2011p353} studied the relations between spectral lines and {\lc} and host-galaxy properties using different spectroscopic SDSS samples. The full three-year sample was used by \citet{Lampeitl:2010p317} to analyze the effect of global host-galaxy properties on {\lc} parameters; \citet{Smith:2011p8634} studied the SN Ia rate as a function of host-galaxy properties; \citet{2011ApJ...743..172D} correlated the Hubble residuals of SNe Ia to the global star-formation rate in their host galaxies, and \citet{Gupta:2011p4678} related the ages and masses of the SN Ia host galaxies to SN properties.

In this analysis we restrict the sample to redshifts $z<0.25$, where the detection efficiency of the SDSS-II SN survey remains reasonably high ($\gtrsim 0.5$, \citealp{Smith:2011p8634}).
This constraint provides a sample of 608 SNe Ia, of which 376 have been confirmed spectroscopically and 232 are photometrically classified SNe Ia.
\subsection{Host Galaxy Identification}
We have matched every SN Ia in our sample to the closest galaxy within an angular separation of 20\arcsec~using the SDSS Data Release 7 (DR7) data set \citep{Abazajian:2009p425}, which contains imaging of more than 8\,000 deg$^2$ of the sky in the five SDSS optical bandpasses $ugriz$ \citep{Fukugita:1996p522} including the 300~deg$^2$ of Stripe 82 where the SDSS-II SN sample is located. The matching was done through the SDSS Image Query Form\footnote{\href{http://cas.sdss.org/astrodr7/en/tools/search/IQS.asp}{http://cas.sdss.org/astrodr7/en/tools/search/IQS.asp}}. Out of the 608 SNe in the redshift range of this analysis, 17 SNe did not have a visible galaxy within 20\arcsec~ and were consequently excluded from the following analysis, leaving 591 SNe Ia (363 Spec-Ia and 228 Photo-Ia).

\section{Measurements}
\subsection{{\lccap} Parameters}
\label{sec:lcparam}
We fit the SN Ia {\lc}s with two {\lc} fitters ({\mlcs} and {\salttwo}) using the implementation in the publicly available Supernova Analyzer package (SNANA\footnote{We used version 9.41 of SNANA, available at \href{http://sdssdp62.fnal.gov/sdsssn/SNANA-PUBLIC/}{http://sdssdp62.fnal.gov/sdsssn/SNANA-PUBLIC/}}, \citet{Kessler:2009p306}).

For the {\mlcs} fitter we use $R_{V} = 2.2$ for the reddening law and an $A_{V}$ prior of $P(A_{V}) ={\rm exp}(-A_{V} /\tau)$ with $\tau = 0.33$, as described in \citet{Kessler:2009p306}. We have checked that using $R_V=3.1$ instead does not qualitatively change the results.
The fitter provides four parameters for each SN: epoch of maximum brightness ($t_{0}$), {\lc} extinction ($A_{V}$), decline rate of the {\lc} ($\Delta$), and distance modulus ($\mu_{MLCS}$).

In the {\salttwo} {\lc} fitter the epoch of maximum brightness ($t_{0}$), the color variability of the supernova ($c$), the stretch of the {\lc} ($x_1$), and the apparent magnitude at maximum brightness in the $B$ band ($m_{B}$) are determined from the fit to the {\lc}. The distance modulus can be calculated by
\begin{equation}
\mu_{SALT2}= m_{B}-M +\alpha x_1 - \beta c \, ,
\end{equation}
where $M$, $\alpha$ and $\beta$ are obtained by minimizing the Hubble diagram residuals. For the average absolute magnitude at peak brightness ($M$) we use  $-19.41\pm0.04$ (\citealt{Guy:2005p86}, where $H_{0}=70~{\rm km~s^{-1}~Mpc^{-1}}$ was used\footnote{The value of the Hubble constant does not affect the results of this analysis.}). For $\alpha$ and $\beta$ we use the values obtained from the three-year SDSS-II SN sample, independent of cosmology ($\alpha=0.135\pm0.033$ and $\beta=3.19\pm0.24$, \citet{Marriner:2011p230}). 

Both $\Delta$ and $x_1$ are related to the width of the supernova {\lc}. However, while $\Delta$ increases for narrower {\lc}s, $x_1$ decreases.
The $A_{V}$ and $c$ parameters are both measurements of color variability. {\mlcs} assumes that the color variations not included in $\Delta$ can be described by a Milky Way-like dust extinction law with an unknown total-to-selective extinction ratio (usually denoted $R_V$) constant for the full sample and an $A_{V}$ that varies between individual supernovae. The $c$ parameter in {\salttwo} describes the color variation of a SN Ia relative to a fiducial SN Ia model, and it includes both the extinction by dust in the host galaxy and the intrinsic color variation independent of $x_1$.

The Hubble residual is defined as $\delta\mu_{{\rm{fit}}} \equiv \mu_{{\rm{fit}}}-\mu_{{\rm{cosmo}}}$ where $_{{\rm{fit}}}$ is either $_{\rm{MLCS}}$ or $_{\rm{\salttwo}}$, depending on the {\lc} fitter, and
\begin{equation}
\mu_{{\rm{cosmo}}}=25+5~{\rm log}_{10}\left[\frac{c}{H_{0}}(1+z_{\rm{SN}}) \int_{0}^{z_{\rm{SN}}} \frac{dz'}{\sqrt{\Omega_{M}(1+z')^{3}+\Omega_{\Lambda}}}\right]
\end{equation}
is the distance modulus calculated using the supernova redshift ($z_{\rm{SN}}$) and a fiducial cosmology. We assume a flat cosmology with $\Omega_{M}=0.274=1-\Omega_{\Lambda}$, and the value for $H_{0}$ which minimizes the scatter for each sample.
The uncertainties in the Hubble residuals were taken as the uncertainties in the distance moduli extracted from the {\lc} fit, $\mu_{\rm{SALT2}}$ and $\mu_{\rm{MLCS2k2}}$ respectively, without adding any contributions from a possible intrinsic dispersion in the distance moduli.
\subsubsection{{\lccap} selection cuts}
To assure robust {\lc} parameters, we applied similar selection cuts as in the SDSS-II SN first year cosmology paper \citep{Kessler:2009p310}.
For {\mlcs}  ({\salttwo}), we used the following requirements:
\begin{itemize}
\item At least 5 photometric observations at different epochs between $-20$ and $+70$ days ($+60$ days for {\salttwo}) in the supernova rest frame relative to peak brightness in $B$ band.
\item At least one measurement earlier than 2 days (0 days) in the rest frame before the date of $B$-band maximum.
\item At least one measurement later than 10 days (9.5 days) in the rest frame after the date of $B$-band maximum.
\item At least one measurement with a signal-to-noise ratio greater than 5 for each of the $g, r$ and $i$ bands (not necessarily from the same night).
\item A {\lc} fit probability of being a SN Ia, based on the $\chi^2$ per degree of freedom, greater than 0.001. 
\end{itemize}
These cuts were designed to remove objects with questionable classification, uncertain determination of the time of maximum brightness, or peculiar or badly constrained {\lc}s. 

Out of the 591 objects, there are 248 that fail the selection cuts for {\mlcs}, leaving 343 SNe (228 Spec-Ia and 115 Photo-Ia). For {\salttwo}, there are 249 objects that fail, leaving 342 SNe (217 Spec-Ia and 125 Photo-Ia). Note that the {\mlcs} and {\salttwo} samples are studied separately. The majority of the remaining SNe are present in both samples, but some are retained in one but not the other.

Furthermore, we remove all SNe with extreme values of the {\lc} parameters, in order to have a sample unaffected by peculiar objects. 
We follow the empirically determined cuts in~\citet{Lampeitl:2010p317} which define the location in the {\lc} parameter space for the majority of SNe Ia in the SDSS-II SN sample. For {\mlcs} we restrict the sample to  $\Delta>-0.4$, removing 30 SNe, while for {\salttwo} the allowed ranges are set to $-0.3<c<0.6$ and $-4.5<x_1<2.0$, removing 22 SNe. After the cuts on {\lc} parameters, 313 SNe (203 Spec-Ia and 110 Photo-Ia) remain in the {\mlcs} sample, and 320 (209 Spec-Ia and 111 Photo-Ia) in the {\salttwo} sample. 

\subsection{Host Galaxy Typing}
\label{sec:galaxytyping}
We split the supernova sample into two groups depending on the morphology of the host galaxy determined using two photometric parameters: the inverse concentration index, and the comparison of the likelihoods for two different S\'ersic brightness profiles \citep{Sersic:1963p2228}. 

The inverse concentration index \citep[e.g.][]{Strateva:2001p564,Shimasaku:2001p488} is defined as the ratio between the radii of two circles, centered on the core of the galaxy, containing respectively $50\%$ and $90\%$ of the Petrosian flux (see \citet{Blanton:2001p8617}). These radii are obtained in the $r$ band for all our host galaxies from SDSS DR7 \citep{Abazajian:2009p425}.

The S\'ersic brightness profile is described by
\begin{equation}
I(r)=I_{0}~{\rm exp}\left[-a_{n}\left( \frac{r}{r_{e}} \right)^{1/n}\right] \ ,
\end{equation}
where $r$ is the distance from the galaxy center, $I_{0}$ is the intensity at the center ($r = 0$) and $r_{e}$ is the radius which contains half of the luminosity. From SDSS DR7 we obtain the $r$-band profiles of all host galaxies for two specific patterns: a pure exponential profile ($n=1$, $a_{1}=1.68$) and a de Vaucouleurs profile ($n=4$, $a_{4}=7.67$) \citep[see][and references therein]{Ciotti:1991p2238, Graham:2005p256}. We also extract from SDSS DR7 the likelihoods for the two fits. The exponential profile is better at describing the decrease in brightness for spiral galaxies, while the de Vaucouleurs profile is better at describing elliptical galaxies \citep{deVaucouleurs:1948p89,Freeman:1970p393}. 

\begin{figure} [!ht]
\begin{center}
\includegraphics[width=\textwidth]{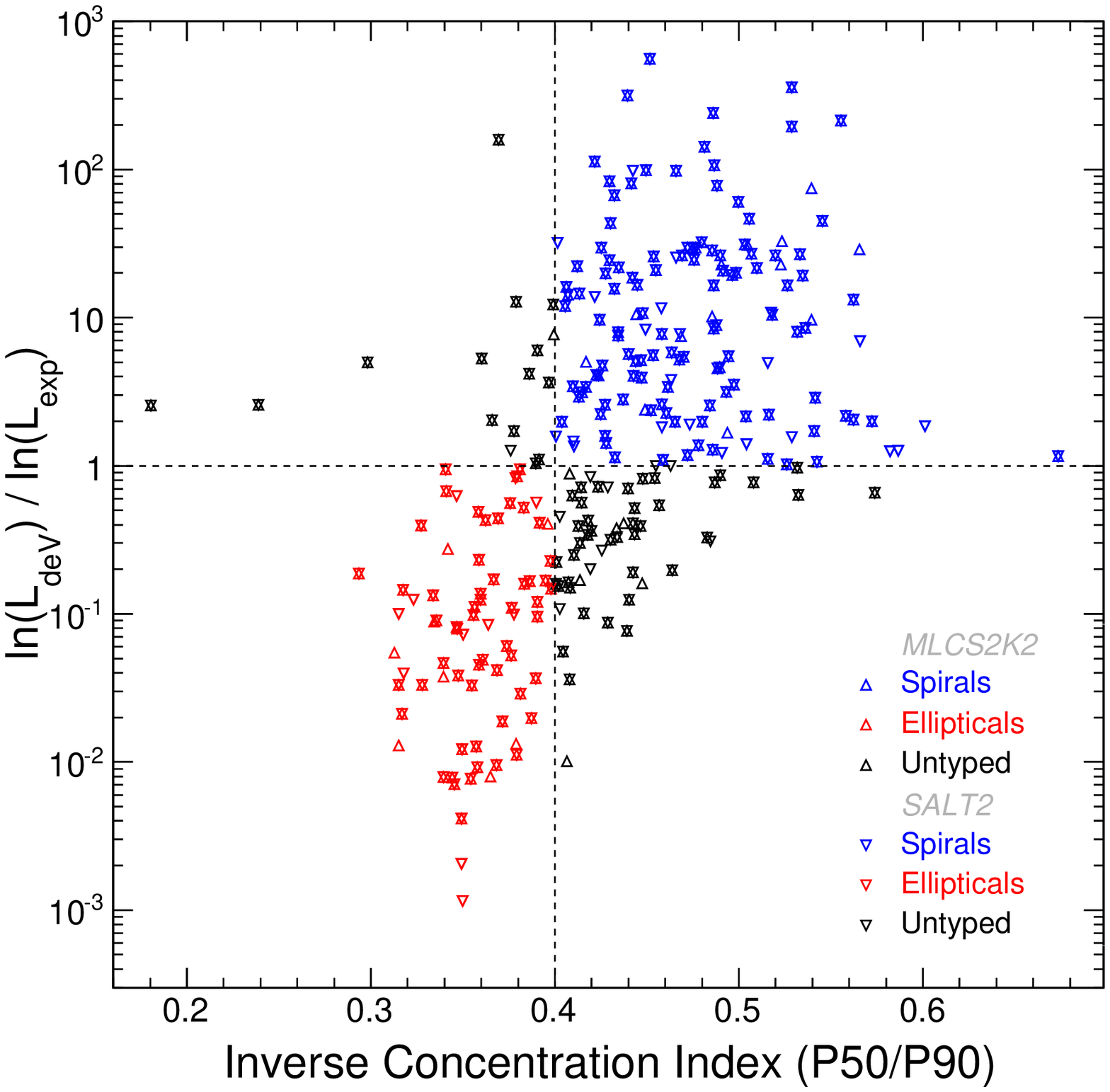}
\end{center}
\caption{Determination of the morphology of the host galaxies using the inverse concentration index and the comparison of the likelihoods for the fits to a de Vaucouleurs and an exponential S\'ersic brightness profile. The vertical axis shows the ratio of the logarithmic likelihoods. The dashed lines show the separation points between elliptical and spiral galaxies. The two methods must agree in order for a galaxy to be classified as either elliptical (red symbols) or spiral (blue symbols). Galaxies with unknown morphology are marked in black. SNe in the {\mlcs} sample are marked with up-pointing triangles, while for SNe in {\salttwo} inverted triangles are used. Those SNe that belong to both samples have the two triangles superimposed.}
\label{fig:typing}
\end{figure}

We assign a elliptical morphology to a galaxy when it has both an inverse concentration index lower than 0.4 \citep{Dilday:2008p396}, and the likelihood for the de Vaucouleurs profile fit is larger than for the exponential fit. A galaxy is classified as a spiral if the inverse concentration index is above 0.4, and the likelihood for the exponential profile fit is larger than for the de Vaucouleurs fit. Figure~\ref{fig:typing} illustrates this separation in morphology. Supernovae for which the two morphological indicators for their host galaxy disagree are removed from the analysis. There are 74 host galaxies which can not be typed within our system as spiral or elliptical galaxies, leaving 239 SNe Ia in the {\mlcs} sample and 246 in the {\salttwo} sample.

%
\subsection{Galactocentric Distances} 
\begin{figure} [!ht]
\begin{center}
\includegraphics[width=12cm]{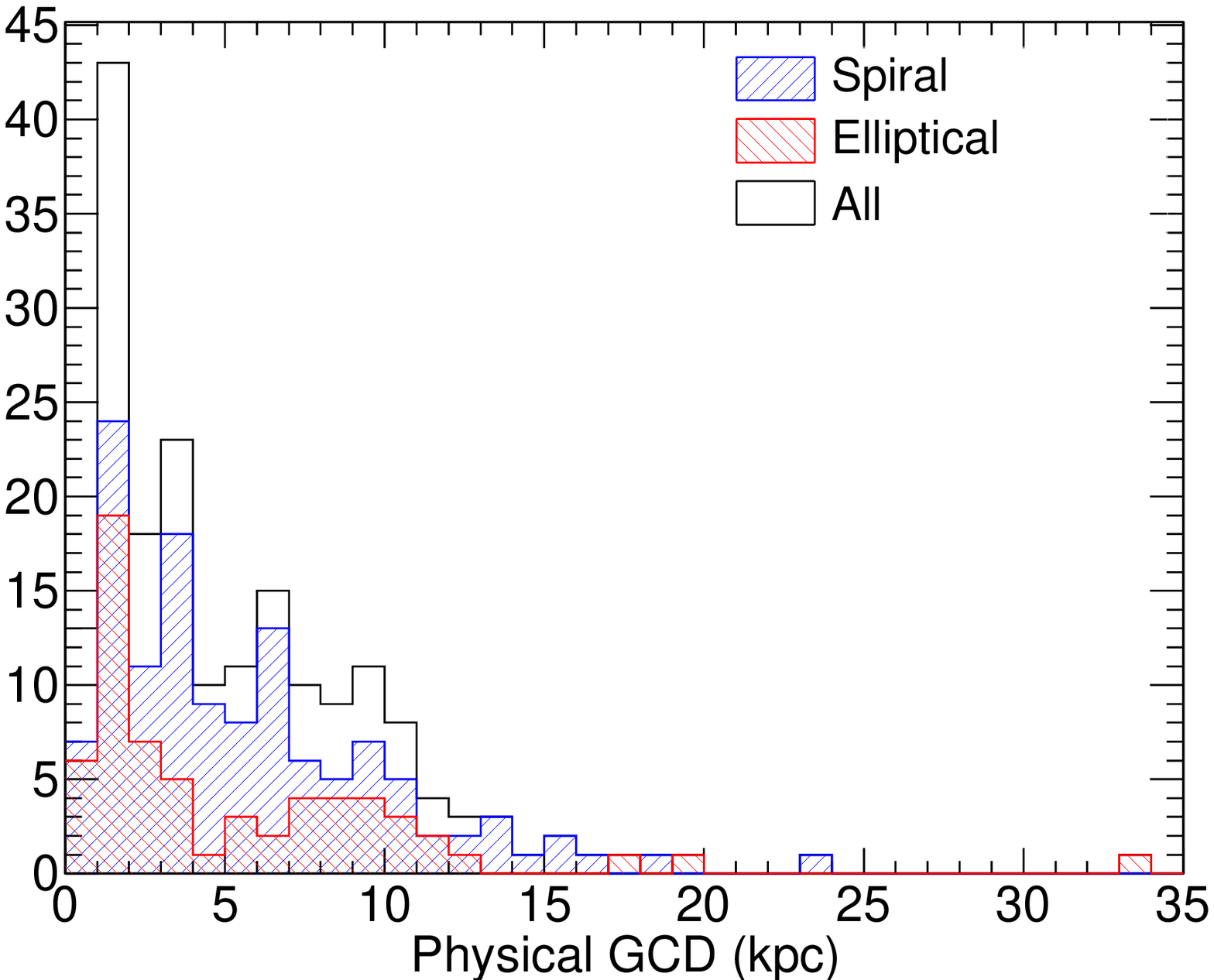}
\includegraphics[width=12cm]{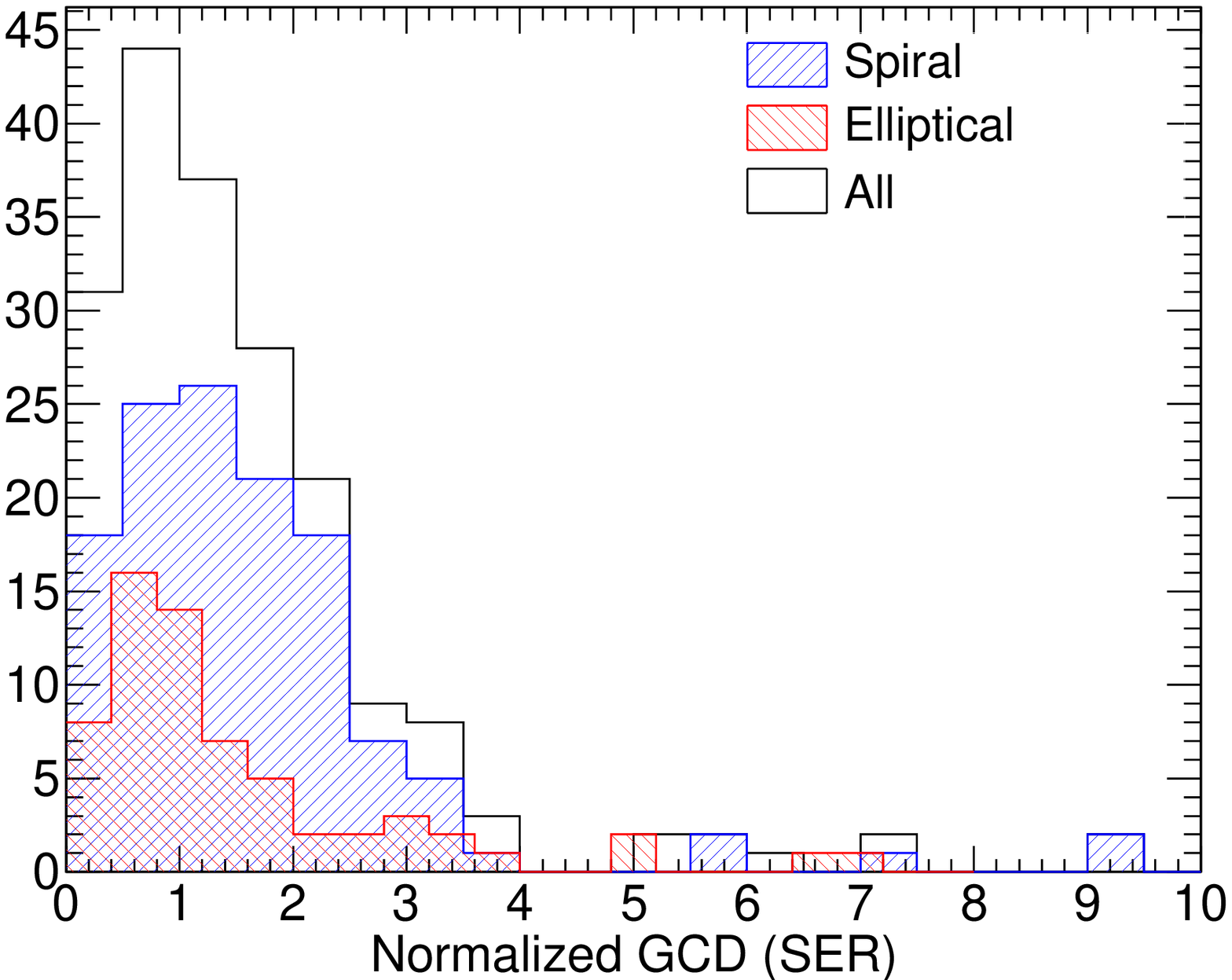}
\end{center}
\caption{Distribution of projected distance between supernova and galaxy core in kiloparsec (top panel) and normalized with the S\'ersic radius (bottom panel) for the SNe Ia present in the final {\mlcs} sample (after all cuts). The sample is divided by the type of the host galaxy. The corresponding figures for {\salttwo} are similar.}
\label{fig:dkpcz}
\end{figure}

From the position of the supernova and the center of the host galaxy, we measure the angular separation between the supernova and its host, and calculate the projected physical distance using the redshift. 
We use the same flat cosmology assumed in the calculation of the Hubble residuals, and a value for the Hubble Constant of $70.4\pm1.4$~km~s$^{-1}$~Mpc$^{-1}$ taken from the Wilkinson Microwave Anisotropy Probe (WMAP) 7-Year results 
The distribution of physical distance of all SNe Ia in our sample is shown in the top panel of Figure~\ref{fig:dkpcz}. \\

Galaxies vary in morphology and size, thus it makes sense to normalize the SN-galaxy separation in order to be able to compare the {\lc} parameters for SNe in the entire host galaxy sample. We use the normalization derived from the shape of the galaxy described by an elliptical S\'ersic profile, taking into account the orientation of the galaxy. We distinguish between elliptical galaxies, which are fitted with a de Vaucouleurs (DEV) profile, and spiral galaxies which are fitted with a pure exponential (EXP) profile (see Section~\ref{sec:galaxytyping} for the definitions). We have repeated the analysis using two other normalizations, one based on the Petrosian 50 radius, defined as the radius of a circle containing $50\%$ of the flux in the $r$ filter, and another using the ellipse estimated from the 25 mag arcsec$^{-2}$ isophote in the $r$ band. The results agree qualitatively with those found using the S\'ersic profile normalization, which are the only ones we will discuss in the following.

The necessary quantities, the major and minor axis and orientation, are obtained from the SDSS DR7 catalogue \citep{Abazajian:2009p425}. The $r$ band is used for all of them. We exclude the supernovae for which any of these quantities is missing.
In the bottom panel of Figure~\ref{fig:dkpcz}, we show the distribution of the normalized distances.

A possible concern may arise from the fact that we use a normalization based on $r$ band galaxy sizes at all redshifts. Thus, since the observer $r$ band will sample bluer restframe wavelengths with increasing redshift, the apparent galaxy sizes, of spiral galaxies in particular, may increase at larger redshifts, resulting in a lower normalized distance. We have found that, indeed, the average spiral galaxy S\'ersic size increases by about 35\% for galaxies above $z=0.1$, compared to lower redshift galaxies, and then it stabilizes beyond $z=0.1$. We have checked that correcting for this has little effect on the correlations with distance that we are  trying to detect, 
and have elected to keep using the uncorrected galaxy sizes obtained from the $r$ band photometry.

All measurements of the distance here are lower limits of the true separation from the center of the host galaxy due to the unknown inclination of the galaxy with respect to the observer. 
We therefore refer to these distances as projected galactocentric distances (GCD). 

We exclude all SNe where the normalized GCD is greater than 10, 
since these SNe are too far from the center of the closest galaxy for the galaxy to be considered as its host with certainty. We also remove all SNe where the normalizing distance (the radius of the galaxy in the direction of the SN) has a large uncertainty: if the radius estimate has a fractional error larger than 100~\% or an absolute error larger than 0.5\arcsec. We also apply a cut on the SN-galaxy distance if the uncertainty in the distance is larger than the actual distance, or 
if the uncertainty in the distance is larger than either 0.5\arcsec~or 1~kpc. The cuts were motivated by the distribution of errors for the full sample.

There are 49 SNe in the {\mlcs} sample and 51 in the {\salttwo} sample which are excluded from the analysis because the matched host galaxy lack one or more of the parameters needed for the distance calculation, because the supernova is too far from the center of the matched host, or because the uncertainty in the galaxy size or galaxy-supernova distance is too large.

\begin{figure} [!ht]
\begin{center}
\includegraphics[width=9cm]{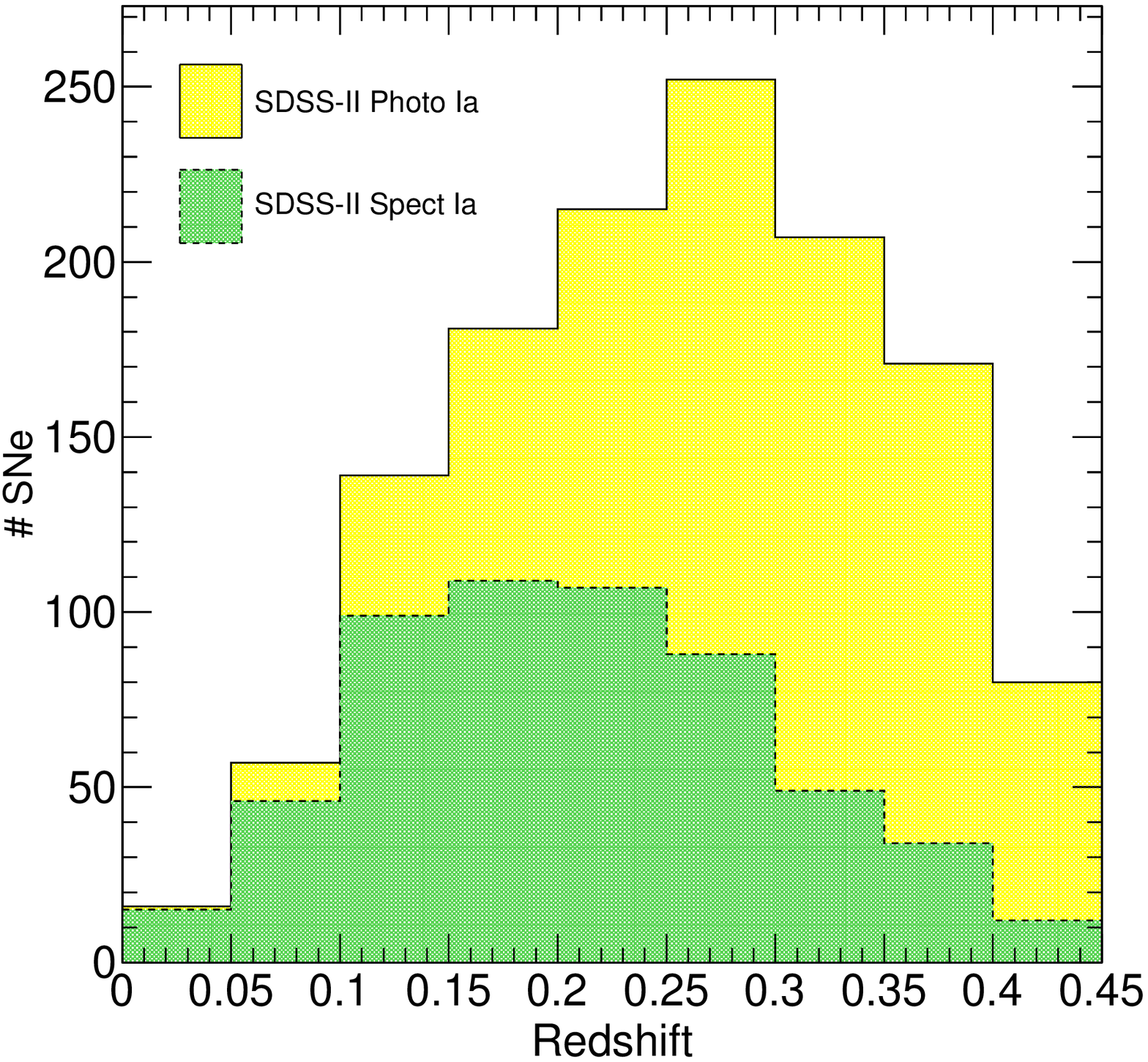}
\includegraphics[width=9cm]{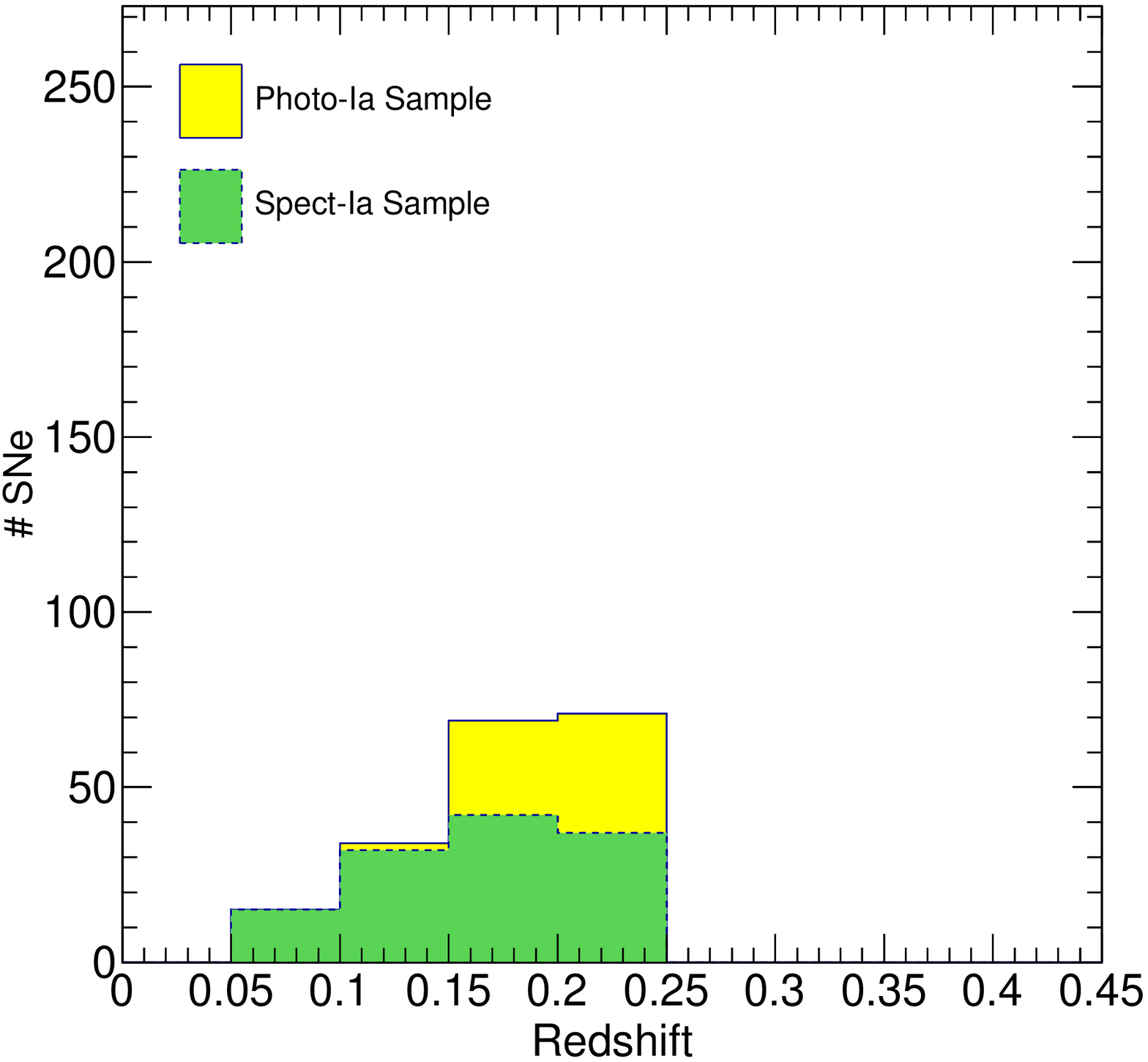}
\end{center}
\caption{Redshift distribution for the 1318 SNe Ia in the full SDSS-II SN sample ($z < 0.45$) (top panel) and for the sample used in this analysis after all cuts have been applied (bottom panel), divided into spectroscopically confirmed SNe Ia and photometrically identified SNe Ia. The bottom panel shows the 190 SNe Ia in the sample used with the {\mlcs} fitter. The corresponding figure for the {\salttwo} fitter is similar.}
\label{fig:sample}
\end{figure}

\begin{table}[h]
\caption{Number of SNe Ia in the sample after applying various selection cuts.}
\begin{center}
\begin{tabular}{lcccccccc}
\hline \hline
 & \multicolumn{2}{c}{Spec-Ia} & &\multicolumn{2}{c}{Photo-Ia} & &\multicolumn{2}{c}{Total}\\ \cline{2-3} \cline{5-6}  \cline{8-9}
 & {\mlcsshort} & {\salttwo} & & {\mlcsshort} & {\salttwo} & & {\mlcsshort} & {\salttwo}\\
 \hline
SN Ia sample  ($z<0.45$)  & \multicolumn{2}{c}{559} & &\multicolumn{2}{c}{759} & & \multicolumn{2}{c}{1318}\\
 \hline
Redshift $<0.25$ & \multicolumn{2}{c}{376} & & \multicolumn{2}{c}{232} & & \multicolumn{2}{c}{608} \\
Identified host galaxy & \multicolumn{2}{c}{363} & & \multicolumn{2}{c}{228} & & \multicolumn{2}{c}{591} \\
LC quality cuts & 228 & 217 & &115 & 125 & & 343 & 342\\ 
LC parameter cuts  & 203 & 209 & & 110 & 111 & & 313 & 320\\
Determined host type  & 160 & 164 & & 79 & 82 & & 239 & 246\\ 
Distance cuts & 127 & 131 & & 63 & 64 & & 190 & 195\\
\hline \hline
\end{tabular}
\end{center}
\label{tab:sample}
\end{table}
Finally, after all cuts are applied, the analysis of the {\lc} parameters as a function of the separation to the center of the supernova host is performed with 190 SNe for {\mlcs} and 195 for {\salttwo}. In Table~\ref{tab:sample} we present the number of SNe before and after each selection cut. 
The list of all SNe used in this analysis is given in Table~\ref{mlcstable}, where we indicate if the SN is present only in the {\mlcs} or {\salttwo} samples or in both. The redshift, the estimated galactocentric distances and host type will be released in \citealp{Sako:2012}, together with all the SDSS-II SN sample data, and the SDSS-III SN redshifts will be released in \citealp{Olmstead:2012}.
In Fig.~\ref{fig:sample} the redshift distribution of the SNe is shown. The final sample consists of 64 SNe in elliptical host galaxies and 126 SNe in spiral galaxies for the {\mlcs} sample. For the {\salttwo} sample there are 65 SNe in elliptical galaxies and 130 in spiral galaxies.

\begin{table}[h]\tiny
\caption{Final SN sample after all cuts have been applied. \textcolor{black}{There remain 190 SNe for MLCS, and 195 for SALT2.}} 
\begin{center}
\begin{tabular}{cc|cc|cc}
\hline \hline 
ID${}^a$ & 
In sample${}^b$ & 
ID${}^a$ & 
In sample${}^b$ & 
ID${}^a$ & 
In sample${}^b$  \\
\hline
2004hz	&	both	&	2006ku	&	both	&	2007ro	&	{\salttwo}	\\
2004ie	&	both	&	2006kw	&	both	&	2007sb	&	both	\\
2005eg	&	both	&	2006kx	&	both	&	779	&	both	\\
2005ez	&	both	&	2006ky	&	{\salttwo}	&	911	&	both	\\
2005fa	&	{\salttwo}	&	2006kz	&	both	&	1415	&	both	\\
2005ff	&	both	&	2006la	&	both	&	2057	&	both	\\
2005fm	&	both	&	2006lb	&	{\salttwo}	&	2162	&	both	\\
2005fp	&	both	&	2006lj	&	both	&	2639	&	both	\\
2005fu	&	both	&	2006lo	&	both	&	3049	&	both	\\
2005fv	&	both	&	2006lp	&	both	&	3426	&	both	\\
2005fw	&	both	&	2006md	&	both	&	3959	&	both	\\
2005fy	&	{\mlcs}	&	2006mt	&	both	&	4019	&	both	\\
2005ga	&	both	&	2006mv	&	both	&	4690	&	{\mlcs}	\\
2005gb	&	both	&	2006mz	&	both	&	5199	&	both	\\
2005gc	&	both	&	2006nb	&	both	&	5486	&	both	\\
2005gd	&	both	&	2006nc	&	both	&	5689	&	both	\\
2005ge	&	both	&	2006ni	&	both	&	5785	&	{\salttwo}	\\
2005gf	&	both	&	2006nn	&	{\salttwo}	&	5859	&	both	\\
2005gp	&	both	&	2006no	&	both	&	5963	&	both	\\
2005gx	&	both	&	2006od	&	both	&	6274	&	both	\\
2005hj	&	both	&	2006of	&	{\salttwo}	&	6326	&	both	\\
2005hn	&	both	&	2006oy	&	{\mlcs}	&	6530	&	both	\\
2005hv	&	both	&	2006pa	&	{\salttwo}	&	6614	&	both	\\
2005hx	&	both	&	2007hx	&	both	&	6831	&	both	\\
2005hy	&	{\salttwo}	&	2007ih	&	both	&	6861	&	both	\\
2005hz	&	both	&	2007ik	&	both	&	7350	&	{\mlcs}	\\
2005if	&	both	&	2007jd	&	{\mlcs}	&	7600	&	both	\\
2005ij	&	both	&	2007jk	&	both	&	8254	&	both	\\
2005ir	&	both	&	2007jt	&	both	&	8555	&	both	\\
2005is	&	both	&	2007ju	&	both	&	9740	&	both	\\
2005je	&	{\salttwo}	&	2007jw	&	{\mlcs}	&	9817	&	{\mlcs}	\\
2005jh	&	both	&	2007jz	&	both	&	10106	&	{\mlcs}	\\
2005jk	&	both	&	2007kb	&	both	&	11172	&	{\salttwo}	\\
2005jl	&	both	&	2007kq	&	both	&	12804	&	both	\\
2005js	&	{\mlcs}	&	2007ks	&	both	&	13323	&	both	\\
2005kp	&	both	&	2007kt	&	both	&	13545	&	both	\\
2005kt	&	both	&	2007kx	&	both	&	13897	&	both	\\
2005mi	&	{\salttwo}	&	2007lc	&	both	&	13907	&	both	\\
2006er	&	both	&	2007lg	&	both	&	14113	&	{\salttwo}	\\
2006ex	&	both	&	2007li	&	{\salttwo}	&	14317	&	both	\\
2006ey	&	both	&	2007lk	&	{\mlcs}	&	14389	&	both	\\
2006fa	&	both	&	2007lo	&	both	&	14445	&	{\salttwo}	\\
2006fb	&	both	&	2007lp	&	both	&	14525	&	both	\\
2006fc	&	{\mlcs}	&	2007lq	&	both	&	14554	&	{\mlcs}	\\
2006fl	&	both	&	2007lr	&	{\salttwo}	&	14784	&	both	\\
2006fu	&	both	&	2007ly	&	{\mlcs}	&	15033	&	both	\\
2006fx	&	both	&	2007ma	&	both	&	15343	&	both	\\
2006fy	&	{\mlcs}	&	2007mb	&	both	&	15587	&	both	\\
2006gg	&	both	&	2007mc	&	both	&	15748	&	both	\\
2006gp	&	{\salttwo}	&	2007mh	&	both	&	15823	&	both	\\
2006gx	&	{\mlcs}	&	2007mi	&	{\salttwo}	&	15829	&	both	\\
2006he	&	{\salttwo}	&	2007mj	&	both	&	15850	&	both	\\
2006hh	&	both	&	2007mz	&	{\salttwo}	&	15866	&	both	\\
2006hl	&	both	&	2007ne	&	both	&	16052	&	both	\\
2006hp	&	both	&	2007nf	&	both	&	16103	&	both	\\
2006hr	&	{\mlcs}	&	2007ni	&	{\salttwo}	&	16163	&	both	\\
2006hw	&	both	&	2007nj	&	both	&	16452	&	{\salttwo}	\\
2006iy	&	both	&	2007nt	&	both	&	16462	&	both	\\
2006ja	&	both	&	2007oj	&	both	&	16467	&	both	\\
2006jn	&	both	&	2007ok	&	both	&	17206	&	both	\\
2006jp	&	{\mlcs}	&	2007om	&	both	&	17408	&	{\salttwo}	\\
2006jq	&	{\salttwo}	&	2007or	&	{\mlcs}	&	17434	&	both	\\
2006jr	&	both	&	2007ow	&	both	&	17748	&	both	\\
2006jw	&	both	&	2007ox	&	both	&	17908	&	both	\\
2006jy	&	{\salttwo}	&	2007oy	&	both	&	17928	&	both	\\
2006jz	&	both	&	2007pc	&	both	&	18362	&	both	\\
2006ka	&	both	&	2007pt	&	both	&	19317	&	{\mlcs}	\\
2006kd	&	both	&	2007qf	&	{\salttwo}	&	19987	&	both	\\
2006kl	&	both	&	2007qh	&	both	&	20088	&	{\mlcs}	\\
2006kq	&	{\mlcs}	&	2007qo	&	both	&	20232	&	both	\\
2006ks	&	both	&	2007qq	&	both	&	20480	&	both	\\
2006kt	&	both	&	2007rk	&	both	&	20721	&	both	\\
 \hline \hline
\multicolumn{6}{l}{${}^a$ IAU name when exists, otherwise internal SDSS name.}\\
\multicolumn{6}{l}{${}^{b}$ Indicates if SN is present only in the {\mlcs} or {\salttwo} samples, or in both.} \\
\end{tabular}
\end{center}
\label{mlcstable}
\end{table}

%
\section{Results}
We have searched for trends in SN Ia {\lc} parameters with GCD. The photometric and the spectroscopic sub-samples were analysed together since 
no significant differences were detected 
between them. The results obtained hold for both sub samples. We examined correlations for the complete sample, as well as when dividing the sample according to host galaxy morphology (spiral and elliptical).

\begin{figure} [!ht]
\begin{center}
\includegraphics[width=15cm]{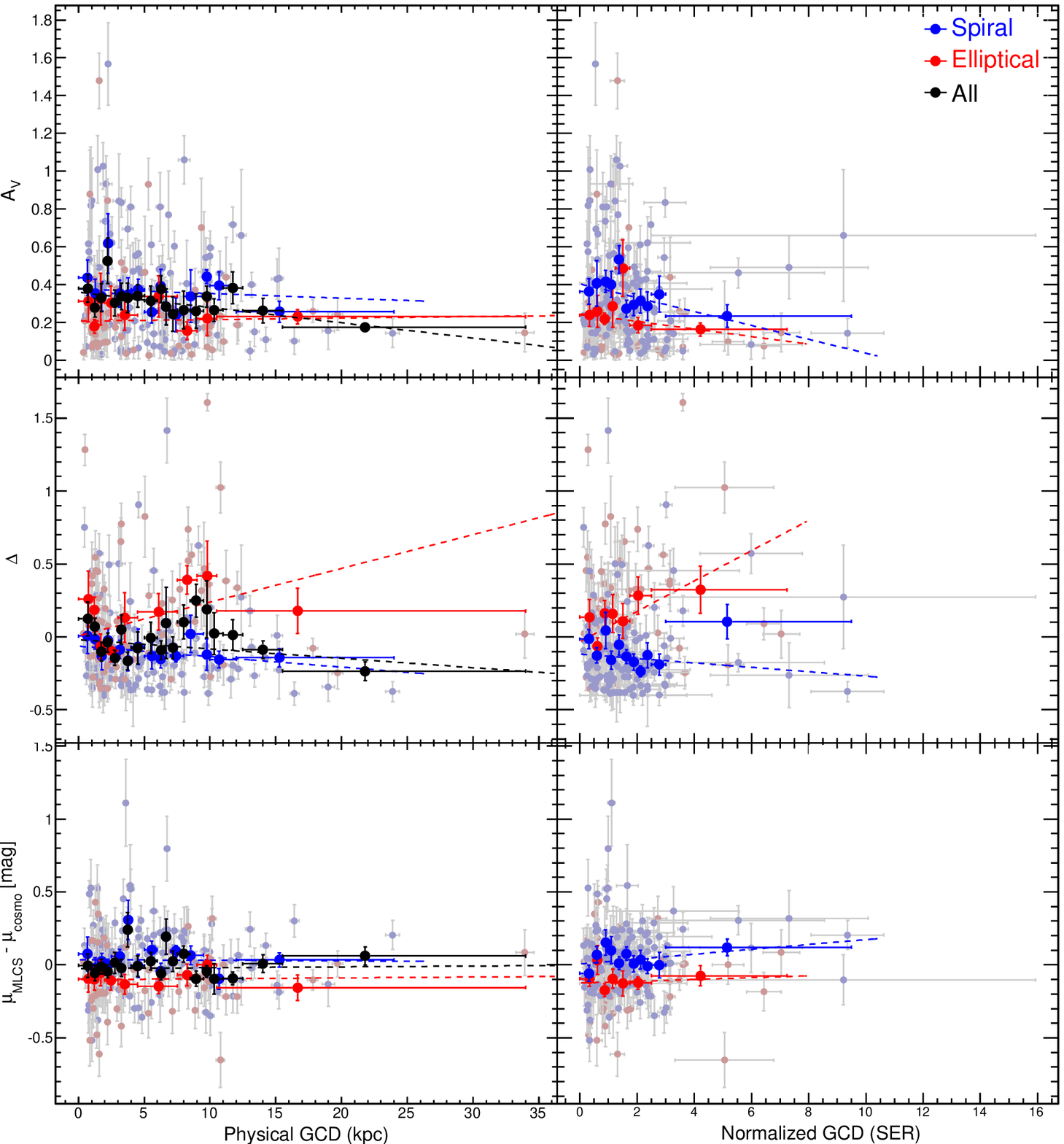}
\end{center}
\caption{{\mlcs} parameters and Hubble residuals as a function of projected distance in kiloparsec and S\'ersic normalization.  SNe in elliptical galaxies are marked in red and SNe in spiral galaxies in blue. Each individual supernova is shown as a small dot, and the bold points indicate the mean values in each bin. Note that the error bars in distance for the binned data show the extent of the bin, and not the standard deviation of the points. The dotted lines show the best fit to the mean values. The values for spiral galaxies and elliptical galaxies in the plots for the S\'ersic profile cannot be directly compared since they have different normalizations (EXP and DEV) and thus there is no black line showing the combined result.}
\label{fig:jhamlcs}
\end{figure}

\begin{figure} [!ht]
\begin{center}
\includegraphics[width=15cm]{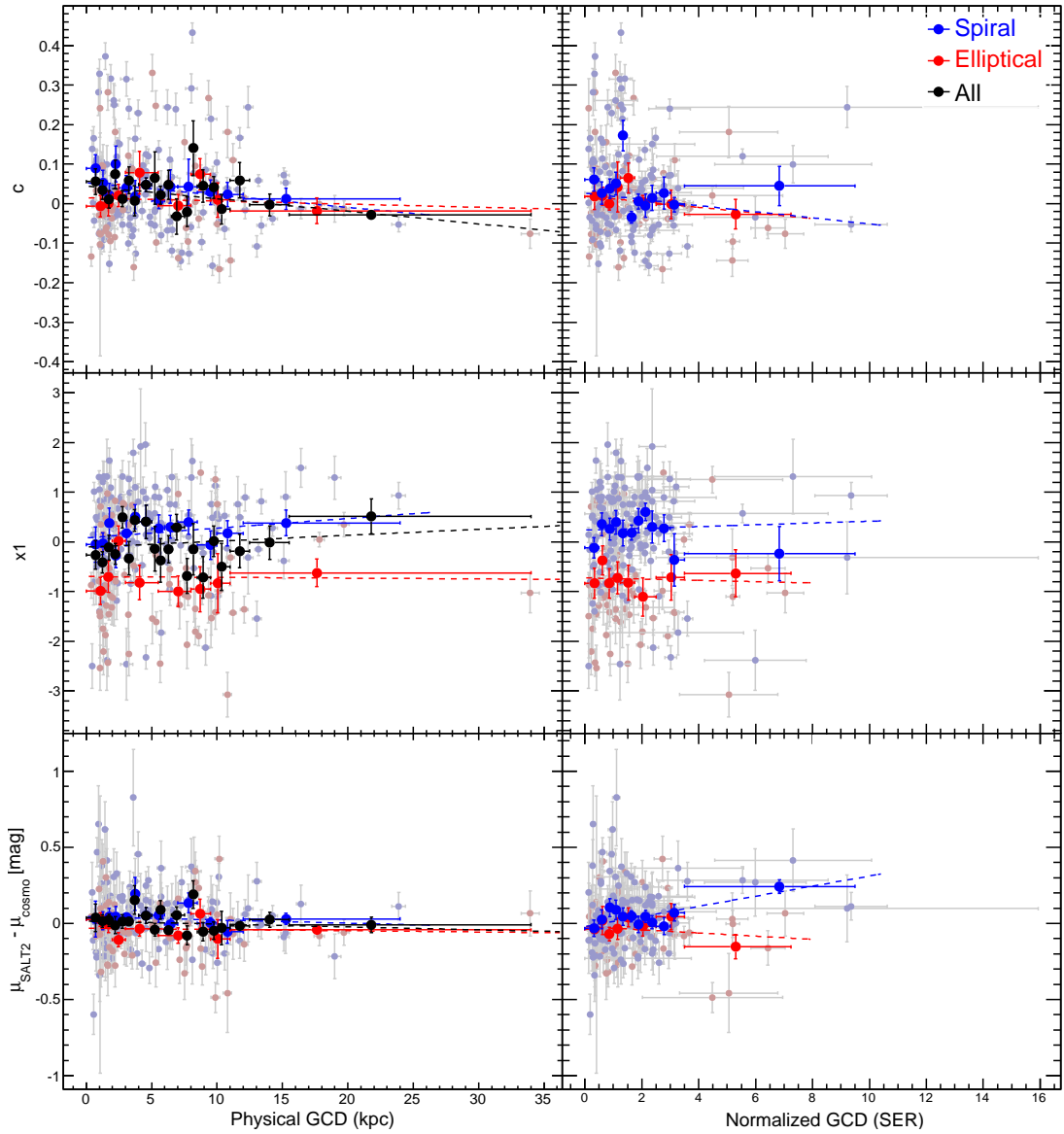}
\end{center}
\caption{{\salttwo} parameters and Hubble residuals as a function of projected distance in kiloparsec and S\'ersic normalization. The same format is used as for Figure~\ref{fig:jhamlcs}.}
\label{fig:jhasalt2}
\end{figure}

We correlate four {\lc} parameters ({\mlcs}: $A_{V}, \Delta$ and {\salttwo}: $x_1, c$) and the Hubble residuals with two different measurements of the distance to the center of the host galaxy (physical GCD, and normalized GCD expressed in S\'ersic (DEV/EXP) radius). For every combination of {\lc} parameter and distance measurement, we bin the SNe in distance and calculate the mean, both for the {\lc} parameter and the distance. In each bin, the uncertainty in the mean {\lc} parameter is calculated as the RMS in the bin divided by the square root of the number of SNe in the bin. The uncertainty in the distance is taken as the width of the bin.
For the physical GCD, we use a bin width of 0.5~kpc, while for the normalized GCD we use bins of width 0.25. When a bin contains less than 5 SNe, this bin is joined with the next one until there are at least 5 SNe in the bin. We then perform a linear fit to the binned measurements taking into account their uncertainties. The reduced $\chi^2$ is calculated, as well as the significance of the slope (the slope divided by the uncertainty of the slope as obtained from the linear fit).
Figure~\ref{fig:jhamlcs} shows the {\mlcs} parameters for each supernova as a function of the projected separation in kiloparsecs and in S\'ersic units, together with the binned mean values and the best fit lines. Figure~\ref{fig:jhasalt2} shows the corresponding plots for {\salttwo}.
The results from these correlation studies are presented in the upper panels of Tables~\ref{paraAv} to \ref{paraSaltMu}.
For these linear fits to multiple bins, we focus on the results where a dependence with distance is preferred with more than 2$\sigma$ and the reduced $\chi^{2}$ is lower than 2.
A cut in $\chi^2$ is necessary since some of the {\lc} parameters might be correlated with distance, but with a correlation that cannot be modeled with a simple linear fit. For these scenarios we solely study the two-bin analysis (described below), which is model independent.

We also search for the same correlations but using only two bins, ``Near'' and ``Far,'' with equal number of objects in each. Note that this means that the distance where the near/far split is made depends on whether we study all galaxies, spiral galaxies only, or elliptical galaxies only. 
We then calculate the mean values for the two bins, as well as their uncertainties (the RMS of the distribution in the bin divided by the square root of the number of objects in it). We study the significance of the difference in the two means by taking the difference divided by the uncertainty. Finally, we calculate the difference in the scatter for the two bins and compare it with its uncertainty to obtain the significance.
The results from the correlation studies with two bins are presented in the lower panels of Tables~\ref{paraAv} to \ref{paraSaltMu}.
For the two-bin analysis, we focus on results where the difference between the two means or two scatters is greater than 2$\sigma$.

As a cross-check of the fit method, we also fit the measurement points, without binning, with a straight line. The errors on the individual points are increased to include the intrinsic spread in the values, by adding in quadrature a term giving a reduced $\chi^2$ of 1. 

In order to study the effect of non-Ia contamination in the Photo-Ia sample, we repeated the fitting process removing in the Photo-Ia sample all possible combinations of 2 SNe in elliptical hosts and 2 in spiral hosts, which roughly corresponds to the expected 6\% non-Ia contamination. The distribution of the fitted slopes for all the combinations looked consistent with the expectations for no background, within the errors quoted below. Furthermore, we repeated all fits using only the SNe in the Spect-Ia sample (with negligibly small non-Ia contamination), and found the results to be qualitatively similar to the results with the full sample, which are the ones we will present in the following.

We find two (related) trends with high significance and good fit quality: both $A_V$ and $c$ decrease with increasing physical GCD, with the slopes of the linear fits being respectively 4.8 and 4.4~$\sigma$ away from zero. These and other correlations with lower significance are presented in detail in the following. 


\subsection{Correlations between projected distance and supernova color ($A_V$, $c$)}
\begin{table}[!ht]\scriptsize
\caption{Results when correlating {\mlcs}-$A_V$ with distance binned in multiple bins of equal size (upper table) and binned in a near and a far sample, with equal number of events in each bin (lower table).} 
\begin{center} 
\begin{tabular}{clcccc} 
\hline \hline 
Distance & Host & Slope & Sig.${}^a$ & $\chi^2$/dof & dof\\ 
unit & type & & & & \\ 
\hline 
\multirow{3}{*}{kpc} & All  & -0.0081 $\pm$ 0.0017 & -4.8 & 0.6 & 17 \\ 
  & Elliptical				& 0.0008 $\pm$ 0.0031 & 0.3 & 0.9 & 7 \\ 
  & Spiral					& -0.0024 $\pm$ 0.0043 & -0.6 & 1.4 & 13 \\ \hline
 deV & Elliptical			& -0.020 $\pm$ 0.012 & -1.6 & 0.8 & 5 \\ \hline
exp & Spiral				& -0.037 $\pm$ 0.015 & -2.4 & 1.2 & 9 \\ \hline
\end{tabular}
\end{center}

\begin{center} 
\begin{adjustwidth}{-0.8in}{-0.8in}
\begin{tabular}{clc|cccc|cccc} 
\hline 
 &  &  & \multicolumn{4}{c}{Mean $A_{V}$} & \multicolumn{4}{c}{Scatter of $A_{V}$}\\ 
\hline 
Distance & Host & Cut${}^b$ & Near & Far & Difference & Sig.${}^a$ & Near & Far & Difference & Sig. ${}^a$\\ 
unit & type & & $\bar{n}$ & $\bar{f}$ & $\bar{f} - \bar{n}$& & $\sigma_n$ & $\sigma_f$ & $\sigma_f - \sigma_n$ & \\ \hline
\hline 
\multirow{3}{*}{kpc}& All & 3.92 & 0.342 $\pm$ 0.031 & 0.305 $\pm$ 0.023 & -0.037 $\pm$ 0.038 & -1.0 & 0.299 $\pm$ 0.037 & 0.224 $\pm$ 0.019 & -0.075 $\pm$ 0.041 & -1.8 \\ 
			& Elliptical      & 3.08 & 0.271 $\pm$ 0.050 & 0.238 $\pm$ 0.036 & -0.033 $\pm$ 0.062 & -0.5 & 0.285 $\pm$ 0.082 & 0.206 $\pm$ 0.039 & -0.079 $\pm$ 0.091 & -0.9 \\ 
			& Spiral          & 4.45 & 0.395 $\pm$ 0.038 & 0.322 $\pm$ 0.028 & -0.073 $\pm$ 0.047 & -1.6 & 0.299 $\pm$ 0.039 & 0.222 $\pm$ 0.022 & -0.078 $\pm$ 0.045 & -1.7 \\ \hline
deV		& Elliptical      & 1.03 & 0.236 $\pm$ 0.033 & 0.274 $\pm$ 0.052 & 0.039 $\pm$ 0.062 & 0.6 & 0.189 $\pm$ 0.044 & 0.295 $\pm$ 0.078 & 0.106 $\pm$ 0.090 & 1.2 \\ \hline
exp		& Spiral          & 1.34 & 0.413 $\pm$ 0.038 & 0.303 $\pm$ 0.027 & -0.110 $\pm$ 0.046 & -2.4 & 0.298 $\pm$ 0.038 & 0.216 $\pm$ 0.022 & -0.082 $\pm$ 0.044 & -1.8 \\ \hline \hline
\multicolumn{8}{l}{} \\\multicolumn{8}{l}{${}^a$ Significance of non-zero result, value divided by uncertainty.} \\\multicolumn{8}{l}{${}^b$ The distance where the `near' and `far' bins were separated.} \\\end{tabular}
\end{adjustwidth}
\end{center}
\label{paraAv}
\end{table}

\subsubsection{{\mlcs}}

When studying all SNe Ia, regardless of host galaxy type, we find that the fitted $A_V$ from {\mlcs} decreases with SN-galaxy distance (see Table~\ref{paraAv}).
In the multi-bin analysis we find a deviation from a non-evolving $A_V$ with a 4.8~$\sigma$ significance for physical distances, with a good quality fit (reduced $\chi^2$ of 0.6). 
Using a two-bin analysis, we confirm the sign of the slope, but with lower significance of only 1.0~$\sigma$, due to the loss of precision in using only two bins. Using the linear fit to the unbinned data we find trends of similar significance.

When splitting the sample into SNe in elliptical and spiral galaxies we find indications that the trend of decreasing $A_V$ with distance is driven by the SNe in spiral galaxies, where the deviation from a non-evolving $A_V$ is 0.6 and 2.4~$\sigma$, respectively, when using physical and normalized distances. Similar significances (1.6, 2.4) are found in the two-bin analysis restricted to supernovae in spiral galaxies. This result is also confirmed with the linear fit to the unbinned data.
In all cases, the sample of SNe in elliptical galaxies is consistent with an $A_V$ not evolving with physical distance.

A potentially confusing result from the multi-bin analysis of $A_V$ is that the fit to the full sample, for distances measured in kpc, has a steeper slope than for the samples of SNe in elliptical and spiral galaxies separately. Naively one would expect a slope for the full sample between that of the elliptical and spiral samples. The reason for this seemingly contradictory result is the different binning, e.g., the sample of all SNe has the center of the last bin at a significantly larger distance than the two other samples, thus increasing the lever arm. As a consistency check, we redid the binned analysis, using the same binning for spiral galaxies and the full sample as for the elliptical sample (which is the smallest of the three). Using equal binning, we obtained a fitted line for the full sample which was in between the lines for elliptical and spiral galaxies. We still see a non-zero slope, but with decreased significance because of the lower sensitivity of the fit with fewer bins.
 
Examining Fig.~\ref{fig:jhamlcs}, we can see that the most dimmed explosions are close to the center of their host galaxies. A natural consequence of this result is that the scatter diminishes with distance. This correlation is particularly visible when studying the full set of galaxies, comparing the near and far sub samples split in physical distance (1.8~$\sigma$).
We also find that SNe Ia with high values of $A_V$ preferentially explode in spiral galaxies. Out of the 64 elliptical host galaxies only 6 (9\%) have SNe with an $A_V > 0.5$, while there are 29 (23\%) in the 126 spiral hosts. The mean value of $A_V$ for the SNe in elliptical galaxies was found to be 
$\langle A_V^{\rm{elliptical}} \rangle = 0.26 \pm 0.03$~mag, while for SNe in spiral galaxies it was 
$\langle A_V^{\rm{spiral}}\rangle = 0.36 \pm 0.02$~mag.

\subsubsection{{\salttwo}}

\begin{table}[!ht]\scriptsize
\caption{Results when correlating {\salttwo}-$c$ with distance binned in multiple bins of equal size (upper table) and binned in a near and a far sample, with equal number of events in each bin (lower table).} 
\begin{center} 
\begin{tabular}{clcccc} 
\hline \hline 
Distance & Host & Slope & Sig.${}^a$ & $\chi^2$/dof & dof\\ 
unit & type & & & & \\ 
\hline 
\multirow{3}{*}{kpc} & All  & -0.0032 $\pm$ 0.0007 & -4.4 & 0.9 & 18 \\ 
  & Elliptical				& -0.0008 $\pm$ 0.0022 & -0.4 & 1.0 & 6 \\ 
  & Spiral					& -0.0031 $\pm$ 0.0020 & -1.5 & 0.6 & 11 \\ \hline
deV & Elliptical			& -0.007 $\pm$ 0.008 & -0.9 & 0.5 & 6 \\ \hline
exp & Spiral				& -0.008 $\pm$ 0.007 & -1.2 & 3.4 & 10 \\ \hline
\end{tabular}
\end{center}

\begin{center} 
\begin{adjustwidth}{-0.8in}{-0.8in}
\begin{tabular}{clc|cccc|cccc} 
\hline 
 &  &  & \multicolumn{4}{c}{Mean $c$} & \multicolumn{4}{c}{Scatter of $c$}\\ 
\hline 
Distance & Host & Cut${}^b$ & Near & Far & Difference & Sig.${}^a$ & Near & Far & Difference & Sig. ${}^a$\\ 
unit & type & & $\bar{n}$ & $\bar{f}$ & $\bar{f} - \bar{n}$& & $\sigma_n$ & $\sigma_f$ & $\sigma_f - \sigma_n$ & \\ \hline
\hline 
\multirow{3}{*}{kpc}& All & 3.74 & 0.035 $\pm$ 0.011 & 0.029 $\pm$ 0.012 & -0.006 $\pm$ 0.016 & -0.3 & 0.113 $\pm$ 0.009 & 0.113 $\pm$ 0.010 & 0.000 $\pm$ 0.014 & 0.0 \\ 
			& Elliptical      & 3.27 & 0.009 $\pm$ 0.016 & 0.024 $\pm$ 0.021 & 0.015 $\pm$ 0.027 & 0.6 & 0.093 $\pm$ 0.015 & 0.119 $\pm$ 0.016 & 0.026 $\pm$ 0.022 & 1.2 \\ 
			& Spiral          & 3.94 & 0.047 $\pm$ 0.015 & 0.032 $\pm$ 0.014 & -0.015 $\pm$ 0.020 & -0.7 & 0.119 $\pm$ 0.011 & 0.111 $\pm$ 0.013 & -0.007 $\pm$ 0.017 & -0.4 \\ \hline
deV		& Elliptical      & 1.08 & 0.023 $\pm$ 0.019 & 0.009 $\pm$ 0.019 & -0.013 $\pm$ 0.026 & -0.5 & 0.109 $\pm$ 0.017 & 0.105 $\pm$ 0.014 & -0.004 $\pm$ 0.022 & -0.2 \\ \hline
exp		& Spiral          & 1.31 & 0.059 $\pm$ 0.016 & 0.020 $\pm$ 0.012 & -0.039 $\pm$ 0.020 & -2.0 & 0.126 $\pm$ 0.012 & 0.099 $\pm$ 0.010 & -0.027 $\pm$ 0.016 & -1.7 \\ \hline \hline
\multicolumn{8}{l}{} \\\multicolumn{8}{l}{${}^a$ Significance of non-zero result, value divided by uncertainty.} \\\multicolumn{8}{l}{${}^b$ The distance where the `near' and `far' bins were separated.} \\\end{tabular}
\end{adjustwidth}
\end{center}
\label{paraSaltC}
\end{table}

We now turn to the color term $c$ from the {\salttwo} analysis to determine if we reproduce similar trends (see Table~\ref{paraSaltC}).
For the linear fit to multiple bins, we also find that $c$ decreases, with 4.4~$\sigma$ significance, for the full sample with increasing physical distances.
The corresponding number when only studying spiral galaxies is 1.5~$\sigma$ for the slope with increasing physical distance and 1.2~$\sigma$ with normalized distance. For SNe in elliptical galaxies, the fit is consistent with a non-evolving $c$.

Using the two-bin analysis we confirm the results, but with lower significances, the largest being when using the normalized distance with spiral galaxies (2~$\sigma$). The same result is found when using a linear fit to unbinned data, with significances of similar strengths.

Just as for the {\mlcs} $A_V$ parameter, we find a trend between $c$ and host galaxy type. The mean $c$ for SNe Ia in spiral galaxies is $\langle c^{\rm{spiral}}\rangle = 0.040 \pm 0.010$, while it is $\langle c^{\rm{elliptical}}\rangle = 0.016 \pm 0.013$ for elliptical galaxies.

We find no significant differences between the near and far samples when examining the scatter of the color term $c$.

\subsection{Correlations between projected distance and {\lc} shape ($\Delta$, $x_1$)}

\begin{table}[!ht]\scriptsize
\caption{Results when correlating {\mlcs}-$\Delta$ with distance binned in multiple bins of equal size (upper table) and binned in a near and a far sample, with equal number of events in each bin (lower table).} 
\begin{center} 
\begin{tabular}{clcccc} 
\hline \hline 
Distance & Host & Slope & Sig.${}^a$ & $\chi^2$/dof & dof\\ 
unit & type & & & & \\ 
\hline 
\multirow{3}{*}{kpc} & All  & -0.0064 $\pm$ 0.0031 & -2.1 & 1.8 & 17 \\ 
  & Elliptical				& 0.0231 $\pm$ 0.0092 & 2.5 & 2.4 & 7 \\ 
  & Spiral					& -0.0072 $\pm$ 0.0047 & -1.5 & 0.5 & 13 \\ \hline
deV & Elliptical			& 0.104 $\pm$ 0.043 & 2.4 & 1.3 & 5 \\ \hline
exp & Spiral				& -0.015 $\pm$ 0.021 & -0.7 & 2.1 & 9 \\ \hline
\end{tabular}
\end{center}

\begin{center} 
\begin{adjustwidth}{-0.8in}{-0.8in}
\begin{tabular}{clc|cccc|cccc} 
\hline 
 &  &  & \multicolumn{4}{c}{Mean $\Delta$} & \multicolumn{4}{c}{Scatter of $\Delta$}\\ 
\hline 
Distance & Host & Cut${}^b$ & Near & Far & Difference & Sig.${}^a$ & Near & Far & Difference & Sig. ${}^a$\\ 
unit & type & & $\bar{n}$ & $\bar{f}$ & $\bar{f} - \bar{n}$& & $\sigma_n$ & $\sigma_f$ & $\sigma_f - \sigma_n$ & \\ \hline
\hline 
\multirow{3}{*}{kpc}& All & 3.92 & -0.010 $\pm$ 0.032 & 0.009 $\pm$ 0.039 & 0.019 $\pm$ 0.050 & 0.4 & 0.311 $\pm$ 0.033 & 0.379 $\pm$ 0.048 & 0.069 $\pm$ 0.058 & 1.2 \\ 
			& Elliptical      & 3.08 & 0.067 $\pm$ 0.060 & 0.253 $\pm$ 0.076 & 0.186 $\pm$ 0.096 & 1.9 & 0.337 $\pm$ 0.068 & 0.428 $\pm$ 0.069 & 0.091 $\pm$ 0.097 & 0.9 \\ 
			& Spiral          & 4.45 & -0.077 $\pm$ 0.033 & -0.088 $\pm$ 0.039 & -0.011 $\pm$ 0.051 & -0.2 & 0.262 $\pm$ 0.028 & 0.308 $\pm$ 0.064 & 0.046 $\pm$ 0.070 & 0.7 \\ \hline
deV		& Elliptical      & 1.03 & 0.094 $\pm$ 0.062 & 0.226 $\pm$ 0.075 & 0.132 $\pm$ 0.098 & 1.4 & 0.353 $\pm$ 0.064 & 0.425 $\pm$ 0.073 & 0.072 $\pm$ 0.097 & 0.7 \\ \hline
exp		& Spiral          & 1.34 & -0.059 $\pm$ 0.038 & -0.106 $\pm$ 0.034 & -0.047 $\pm$ 0.051 & -0.9 & 0.302 $\pm$ 0.059 & 0.267 $\pm$ 0.038 & -0.036 $\pm$ 0.070 & -0.5 \\ \hline \hline
\multicolumn{8}{l}{} \\\multicolumn{8}{l}{${}^a$ Significance of non-zero result, value divided by uncertainty.} \\\multicolumn{8}{l}{${}^b$ The distance where the `near' and `far' bins were separated.} \\\end{tabular}
\end{adjustwidth}
\end{center}
\label{paraDelta}
\end{table}

When examining the correlations between the GCD and the {\mlcs} $\Delta$ (Table~\ref{paraDelta}) we find a weak relationship for elliptical galaxies, using the multi-binning method, where larger $\Delta$ are found at larger GCD. The significance of an evolving $\Delta$ is 2.5 and 2.4~$\sigma$ when using physical and normalized distance, respectively. Note that the fit to $\Delta$ as a function of physical distance is of limited quality, with a reduced $\chi^2$ of 2.4. The trend is also visible in the two-bin data, but with lower significance: 1.9 and 1.4~$\sigma$. In the fit to unbinned data the correlation is only seen for normalized distances.

When studying the sample of spiral galaxies, we find only very weak correlations, of the opposite trend as for the SNe in elliptical galaxies. The most significant correlation is with physical distances (1.5~$\sigma$). However, using a two bin analysis and the fit to unbinned data this correlation is even less significant.

Looking at the full sample we see similar trends to what we found for the SNe in spiral galaxies ($\Delta$ diminishes with distance), since there are more SNe in spiral galaxies than in elliptical galaxies in our sample. The trend is visible when studying physical distances in the multi-bin analysis (2.1~$\sigma$), but not in the two-bin analysis.

\begin{table}[!ht]\scriptsize
\caption{Results when correlating {\salttwo}-$x_{1}$ with distance binned in multiple bins of equal size (upper table) and binned in a near and a far sample, with equal number of events in each bin (lower table).} 
\begin{center} 
\begin{tabular}{clcccc} 
\hline \hline 
Distance & Host & Slope & Sig.${}^a$ & $\chi^2$/dof & dof\\ 
unit & type & & & & \\ 
\hline 
\multirow{3}{*}{kpc} & All  & 0.0120 $\pm$ 0.0140 & 0.9 & 1.6 & 18 \\ 
  & Elliptical				& -0.0016 $\pm$ 0.0201 & -0.1 & 1.4 & 6 \\ 
  & Spiral					& 0.0195 $\pm$ 0.0177 & 1.1 & 0.8 & 11 \\ \hline
deV & Elliptical			& -0.013 $\pm$ 0.097 & -0.1 & 0.5 & 6 \\ \hline
exp & Spiral				& 0.018 $\pm$ 0.069 & 0.3 & 1.1 & 10 \\ \hline
\end{tabular}
\end{center}

\begin{center} 
\begin{adjustwidth}{-0.8in}{-0.8in}
\begin{tabular}{clc|cccc|cccc} 
\hline 
 &  &  & \multicolumn{4}{c}{Mean $x_{1}$} & \multicolumn{4}{c}{Scatter of $x_{1}$}\\ 
\hline 
Distance & Host & Cut${}^b$ & Near & Far & Difference & Sig.${}^a$ & Near & Far & Difference & Sig. ${}^a$\\ 
unit & type & & $\bar{n}$ & $\bar{f}$ & $\bar{f} - \bar{n}$& & $\sigma_n$ & $\sigma_f$ & $\sigma_f - \sigma_n$ & \\ \hline
\hline 
\multirow{3}{*}{kpc}& All & 3.74 & -0.164 $\pm$ 0.107 & -0.084 $\pm$ 0.108 & 0.080 $\pm$ 0.152 & 0.5 & 1.059 $\pm$ 0.064 & 1.064 $\pm$ 0.070 & 0.005 $\pm$ 0.095 & 0.1 \\ 
			& Elliptical      & 3.27 & -0.701 $\pm$ 0.181 & -0.830 $\pm$ 0.185 & -0.129 $\pm$ 0.259 & -0.5 & 1.041 $\pm$ 0.089 & 1.049 $\pm$ 0.128 & 0.008 $\pm$ 0.156 & 0.1 \\ 
			& Spiral          & 3.94 & 0.112 $\pm$ 0.118 & 0.280 $\pm$ 0.107 & 0.168 $\pm$ 0.160 & 1.1 & 0.955 $\pm$ 0.092 & 0.866 $\pm$ 0.095 & -0.089 $\pm$ 0.133 & -0.7 \\ \hline
deV		& Elliptical      & 1.08 & -0.775 $\pm$ 0.178 & -0.754 $\pm$ 0.190 & 0.021 $\pm$ 0.260 & 0.1 & 1.021 $\pm$ 0.089 & 1.072 $\pm$ 0.124 & 0.051 $\pm$ 0.153 & 0.3 \\ \hline
exp		& Spiral          & 1.31 & 0.199 $\pm$ 0.114 & 0.192 $\pm$ 0.114 & -0.007 $\pm$ 0.161 & -0.0 & 0.915 $\pm$ 0.092 & 0.915 $\pm$ 0.097 & -0.000 $\pm$ 0.133 & -0.0 \\ \hline \hline
\multicolumn{8}{l}{} \\\multicolumn{8}{l}{${}^a$ Significance of non-zero result, value divided by uncertainty.} \\\multicolumn{8}{l}{${}^b$ The distance where the `near' and `far' bins were separated.} \\\end{tabular}
\end{adjustwidth}
\end{center}
\label{paraSaltX1}
\end{table}

The {\salttwo} $x_1$ parameter provides another measurement of the {\lc} width. Since $x_1$ is inversely proportional to the decline rate of the {\lc} we would expect a correlation with the opposite sign compared to the correlation with {\mlcs}-$\Delta$. We find no correlations with 2~$\sigma$ or larger significance in either the multi-bin or the two-bin analyses (see Table~\ref{paraSaltX1}), the highest significance of a deviation from a constant $x_1$ being only 1.1~$\sigma$. 

Leaving aside the dependence with distance, we confirm the results that faint SNe Ia with narrow {\lc}s favor passive host galaxies \citep{Hamuy:1996p601, Gallagher:2005p1222, Sullivan:2006p504,Lampeitl:2010p317}. We find that SNe Ia with low $\Delta$ / high $x_1$ (bright SNe) explode preferably in spiral galaxies, which is visible in the middle panels of Figures~\ref{fig:jhamlcs}-\ref{fig:jhasalt2}.
We obtain $\langle x_1^{\rm{elliptical}}\rangle = -0.76 \pm 0.13$ for elliptical galaxies compared to $\langle x_1^{\rm{spiral}}\rangle = 0.20 \pm 0.08$ for spiral galaxies. The corresponding numbers for {\mlcs} are: $\langle\Delta^{\rm{elliptical}}\rangle = 0.16 \pm 0.05$~mag and $\langle\Delta^{\rm{spiral}}\rangle = -0.08 \pm 0.03$~mag.

\subsection{Correlations between projected distance and Hubble residuals}
\begin{table}[!ht]\scriptsize
\caption{Results when correlating {\mlcs} Hubble residuals with distance binned in multiple bins of equal size (upper table) and binned in a near and a far sample, with equal number of events in each bin (lower table).} 
\begin{center} 
\begin{tabular}{clcccc} 
\hline \hline 
Distance & Host & Slope & Sig.${}^a$ & $\chi^2$/dof & dof\\ 
unit & type & & & & \\ 
\hline 
\multirow{3}{*}{kpc} & All  & 0.0006 $\pm$ 0.0027 & 0.2 & 1.3 & 17 \\ 
  & Elliptical				& 0.0006 $\pm$ 0.0055 & 0.1 & 0.6 & 7 \\ 
  & Spiral					& -0.0004 $\pm$ 0.0037 & -0.1 & 1.4 & 13 \\ \hline
deV & Elliptical			& 0.006 $\pm$ 0.020 & 0.3 & 0.9 & 5 \\ \hline
exp & Spiral				& 0.017 $\pm$ 0.014 & 1.2 & 0.9 & 9 \\ \hline
\end{tabular}
\end{center}

\begin{center} 
\begin{adjustwidth}{-0.8in}{-0.8in}
\begin{tabular}{clc|cccc|cccc} 
\hline 
 &  &  & \multicolumn{4}{c}{Mean $\delta_{MLCS}$} & \multicolumn{4}{c}{Scatter of $\delta_{MLCS}$}\\ 
\hline 
Distance & Host & Cut${}^b$ & Near & Far & Difference & Sig.${}^a$ & Near & Far & Difference & Sig. ${}^a$\\ 
unit & type & & $\bar{n}$ & $\bar{f}$ & $\bar{f} - \bar{n}$& & $\sigma_n$ & $\sigma_f$ & $\sigma_f - \sigma_n$ & \\ \hline
\hline 
\multirow{3}{*}{kpc}& All & 3.92 & -0.010 $\pm$ 0.026 & 0.007 $\pm$ 0.022 & 0.017 $\pm$ 0.034 & 0.5 & 0.255 $\pm$ 0.029 & 0.217 $\pm$ 0.021 & -0.037 $\pm$ 0.036 & -1.0 \\ 
			& Elliptical      & 3.08 & -0.088 $\pm$ 0.041 & -0.104 $\pm$ 0.033 & -0.016 $\pm$ 0.053 & -0.3 & 0.233 $\pm$ 0.030 & 0.187 $\pm$ 0.030 & -0.046 $\pm$ 0.042 & -1.1 \\ 
			& Spiral          & 4.45 & 0.054 $\pm$ 0.033 & 0.039 $\pm$ 0.025 & -0.015 $\pm$ 0.042 & -0.4 & 0.264 $\pm$ 0.037 & 0.201 $\pm$ 0.025 & -0.063 $\pm$ 0.044 & -1.4 \\ \hline
deV		& Elliptical      & 1.03 & -0.090 $\pm$ 0.039 & -0.102 $\pm$ 0.036 & -0.012 $\pm$ 0.053 & -0.2 & 0.218 $\pm$ 0.028 & 0.205 $\pm$ 0.033 & -0.013 $\pm$ 0.043 & -0.3 \\ \hline
exp		& Spiral          & 1.34 & 0.056 $\pm$ 0.034 & 0.037 $\pm$ 0.024 & -0.018 $\pm$ 0.042 & -0.4 & 0.273 $\pm$ 0.038 & 0.188 $\pm$ 0.015 & -0.085 $\pm$ 0.041 & -2.1 \\ \hline \hline
\multicolumn{8}{l}{} \\\multicolumn{8}{l}{${}^a$ Significance of non-zero result, value divided by uncertainty.} \\\multicolumn{8}{l}{${}^b$ The distance where the `near' and `far' bins were separated.} \\\end{tabular}
\end{adjustwidth}
\end{center}
\label{paraMu}
\end{table}

\begin{table}[!ht]\scriptsize
\caption{Results when correlating {\salttwo} Hubble residuals with distance binned in multiple bins of equal size (upper table) and binned in a near and a far sample, with equal number of events in each bin (lower table).} 
\begin{center} 
\begin{tabular}{clcccc} 
\hline \hline 
Distance & Host & Slope & Sig.${}^a$ & $\chi^2$/dof & dof\\ 
unit & type & & & & \\ 
\hline 
\multirow{3}{*}{kpc} & All  & -0.0019 $\pm$ 0.0022 & -0.9 & 1.0 & 18 \\ 
  & Elliptical				& -0.0008 $\pm$ 0.0024 & -0.3 & 1.1 & 6 \\ 
  & Spiral					& -0.0021 $\pm$ 0.0031 & -0.7 & 0.7 & 11 \\ \hline
deV & Elliptical			& -0.010 $\pm$ 0.016 & -0.7 & 0.8 & 6 \\ \hline
exp & Spiral				& 0.033 $\pm$ 0.008 & 4.0 & 1.0 & 10 \\ \hline
\end{tabular}
\end{center}

\begin{center}
\begin{adjustwidth}{-0.8in}{-0.8in}
\begin{tabular}{clc|cccc|cccc} 
\hline 
 &  &  & \multicolumn{4}{c}{Mean $\delta_{SALT2}$} & \multicolumn{4}{c}{Scatter of $\delta_{SALT2}$}\\ 
\hline 
Distance & Host & Cut${}^b$ & Near & Far & Difference & Sig.${}^a$ & Near & Far & Difference & Sig. ${}^a$\\ 
unit & type & & $\bar{n}$ & $\bar{f}$ & $\bar{f} - \bar{n}$& & $\sigma_n$ & $\sigma_f$ & $\sigma_f - \sigma_n$ & \\ \hline
\hline 
\multirow{3}{*}{kpc}& All & 3.74 & 0.029 $\pm$ 0.022 & 0.009 $\pm$ 0.019 & -0.020 $\pm$ 0.029 & -0.7 & 0.215 $\pm$ 0.022 & 0.183 $\pm$ 0.014 & -0.032 $\pm$ 0.026 & -1.2 \\ 
			& Elliptical      & 3.27 & -0.014 $\pm$ 0.028 & -0.046 $\pm$ 0.037 & -0.032 $\pm$ 0.046 & -0.7 & 0.160 $\pm$ 0.022 & 0.208 $\pm$ 0.027 & 0.048 $\pm$ 0.035 & 1.4 \\ 
			& Spiral          & 3.94 & 0.047 $\pm$ 0.029 & 0.040 $\pm$ 0.020 & -0.007 $\pm$ 0.035 & -0.2 & 0.235 $\pm$ 0.028 & 0.163 $\pm$ 0.014 & -0.071 $\pm$ 0.032 & -2.2 \\ \hline
deV		& Elliptical      & 1.08 & -0.037 $\pm$ 0.028 & -0.022 $\pm$ 0.037 & 0.015 $\pm$ 0.046 & 0.3 & 0.162 $\pm$ 0.023 & 0.207 $\pm$ 0.027 & 0.046 $\pm$ 0.035 & 1.3 \\ \hline
exp		& Spiral          & 1.31 & 0.038 $\pm$ 0.030 & 0.049 $\pm$ 0.019 & 0.011 $\pm$ 0.035 & 0.3 & 0.240 $\pm$ 0.029 & 0.156 $\pm$ 0.012 & -0.084 $\pm$ 0.031 & -2.7 \\ \hline \hline
\multicolumn{8}{l}{} \\\multicolumn{8}{l}{${}^a$ Significance of non-zero result, value divided by uncertainty.} \\\multicolumn{8}{l}{${}^b$ The distance where the `near' and `far' bins were separated.} \\\end{tabular}
\end{adjustwidth}
\end{center}
\label{paraSaltMu}
\end{table}

When correlating the projected distance with the Hubble residuals from {\mlcs} and {\salttwo} (see Table~\ref{paraMu} and \ref{paraSaltMu}), we only find a trend with a significance larger than 2~$\sigma$, indicating that the {\salttwo} Hubble residuals increase with normalized distance for SNe in spiral hosts. However, the trend is only seen in the multi-bin analysis, and it is not confirmed by either the two-bin analysis or the unbinned fit. Furthermore, it is not seen for physical distances, nor in the {\mlcs} residual.

Using the limits obtained from the Hubble residuals as a function of physical distance for the full sample of SNe, we obtain that both the {\mlcs} and {\salttwo} residuals will change by less than 0.06 (2$\sigma$) between a SN at the center of the galaxy and one which is 10 kpc away.

The difference in Hubble residual scatter between SNe in spiral galaxies close to the galaxy center and farther away is significant. Depending on {\lc} fitter and distance type used, the significance varies between 1.4 and 2.7 $\sigma$, which can be seen in Tables~\ref{paraMu} and \ref{paraSaltMu}. The scatter is larger close to the center of the galaxy. This scatter difference translates also to the complete sample, while it is not visible in the elliptical sample only.

Note that we find a difference in Hubble residuals between SNe in spiral galaxies and elliptical galaxies, most notably in the {\mlcs} residuals, 0.05$\pm$0.02~mag in spirals, compared to $-0.10\pm0.03$~mag in ellipticals.

%
\section{Discussion}
Correlating the SN Ia {\lc} parameters with the distance of the supernova from the center of the host galaxies, we find strong indications of a decrease in $A_V$ with distance, in particular for spiral galaxies. 
If part of the color variations of SNe Ia is explained by dust, and dust is mainly present in spiral galaxies and decreasing with distance from the center, this would be expected. The trend is also reproduced when correlating the {\salttwo} color parameter $c$ with distance. We find a moderately significant difference between the mean value of the color parameters $A_V$ and $c$ for SNe exploding in spirals and elliptical galaxies, with  
$\left<A_V\right>_{sp} - \left<A_V\right>_{ell}  = 0.10 \pm 0.04$, and, with less significance, 
$\left<c\right>_{sp}    - \left<c\right>_{ell}       = 0.024 \pm 0.016$. These differences would also be expected if these color parameters are related to dust, more prevalent in spiral galaxies.
Due to the difficulty of observing faint SNe close to the galaxy center, we would expect fewer dust extincted SNe (with high $A_V$) at small distances. However, this is opposite of what we find, so if we corrected for the brightness bias, the trend would most likely be stronger.
Using the first-year SDSS-II/SNe sample, \cite{Yasuda:2010p281} looked for a correlation of the {\salttwo} color parameter $c$ with galactocentric distance, and did not find it significant for distances up to 15~kpc. If we restrict our study to the same region below 15~kpc, we still see a decreasing slope, but now with a smaller significance around 1~$\sigma$. Therefore, our results are consistent with those in \cite{Yasuda:2010p281}, and indicate that the bulk of the effect we see occurs at large distances between the SN and the center of its host galaxy.

We find some indications that SNe in elliptical galaxies tend to have narrower {\lc}s (larger $\Delta$) if they explode farther from the galaxy core. Since the width of the {\lc} is related to the supernova brightness, this result would mean that SNe exploding at larger galactocentric distances are fainter. Therefore, this finding could, at least partly, be explained by the difficulty in detecting faint SNe close to the galaxy center, where the galaxy light is strongest. Furthermore, the significances found for an evolving $\Delta$ are not very strong ($<2.5\sigma$) and the trend is mainly visible when using the $\Delta$ parameter from {\mlcs} as a measure of the {\lc} width, compared to the homologous $x_1$ parameter in {\salttwo}.

We find no strong correlations between the galactocentric distance and the Hubble residuals.
Since the distance of the SN from the core of the galaxy can be used as a proxy for the local metallicity \citep[see e.g.][]{Boissier:2009p192}, this result can be interpreted as an indication of a limited correlation between Hubble residuals and local metallicity. 
Since there is also a correlation between the metallicty and the luminosity of the host galaxy, there could be a bias in our sample where there are fewer SNe detected in bright galaxies (with high metallicity) at small galactocentric distances. However, even if we exclude the data with the smallest SN-galaxy distances, we still see no significant correlations between the galactocentric distance and the Hubble residuals.
\cite{2000ApJ...542..588I} found no correlations between the galactocentric distance and both the absolute magnitude in the $B$ band and the decline rate parameter $\Delta m_{15}$ using 62 SNe at $z<0.1$. Our results, with a larger SN sample that extends to $z<0.25$, agree with those.
\citet{Gallagher:2005p1222} suggest that progenitor age should be a more important factor than metallicity in determining the variations in supernova peak brightness. \citet{Gupta:2011p4678} found a correlation between the Hubble residuals and the mass-weighted average age of the host galaxy in SDSS data. However, a correlation between the Hubble residuals and the \emph{global} metallicity has also been detected \citep{2011ApJ...743..172D}.

\section*{Acknowledgments} 
Funding for the SDSS and SDSS-II has been provided by the Alfred P. Sloan Foundation, the Participating Institutions, the National Science Foundation, the U.S. Department of Energy, the National Aeronautics and Space Administration, the Japanese Monbukagakusho, the Max Planck Society, and the Higher Education Funding Council for England. The SDSS Web Site is http://www.sdss.org/.

The SDSS is managed by the Astrophysical Research Consortium for the Participating Institutions. The Participating Institutions are the American Museum of Natural History, Astrophysical Institute Potsdam, University of Basel, Cambridge University, Case Western Reserve University, University of Chicago, Drexel University, Fermilab, the Institute for Advanced Study, the Japan Participation Group, Johns Hopkins University, the Joint Institute for Nuclear Astrophysics, the Kavli Institute for Particle Astrophysics and Cosmology, the Korean Scientist Group, the Chinese Academy of Sciences (LAMOST), Los Alamos National Laboratory, the Max-Planck-Institute for Astronomy (MPIA), the Max-Planck-Institute for Astrophysics (MPA), New Mexico State University, Ohio State University, University of Pittsburgh, University of Portsmouth, Princeton University, the United States Naval Observatory, and the University of Washington.

This work is based in part on observations made at the following telescopes. 
The Hobby-Eberly Telescope (HET) is a joint project of the University of Texas at Austin, the Pennsylvania State University, Stanford University, Ludwig-Maximillians-Universit\"at M\"unchen, and Georg-August-Universit\"at G\"ottingen. The HET is named in honor of its principal benefactors, William P. Hobby and Robert E. Eberly. The Marcario Low-Resolution Spectrograph is named for Mike Marcario of High Lonesome Optics, who fabricated several optics for the instrument but died before its completion; it is a joint project of the Hobby-Eberly Telescope partnership and the Instituto de Astronom\'ia de la Universidad Nacional Aut\'onoma de M\'exico. 
The Apache Point Observatory 3.5-meter telescope is owned and operated by the Astrophysical Research Consortium. We thank the observatory director, Suzanne Hawley, and site manager, Bruce Gillespie, for their support of this project. 
The Subaru Telescope is operated by the National Astronomical Observatory of Japan. 
The William Herschel Telescope is operated by the Isaac Newton Group, and the Nordic Optical Telescope is operated jointly by Denmark, Finland, Iceland, Norway, and Sweden, both on the island of La Palma in the Spanish Observatorio del Roque de los Muchachos of the Instituto de Astrofisica de Canarias. 
The Italian Telescopio Nazionale Galileo (TNG) is operated on the island of La Palma by the Fundaci\'on Galileo Galilei of the INAF (Istituto Nazionale di Astrofisica) at the Spanish Observatorio del Roque de los Muchachos of the Instituto de Astrof\'isica de Canarias.
Kitt Peak National Observatory, National Optical Astronomy Observatory, is operated by the Association of Universities for Research in Astronomy, Inc. (AURA) under cooperative agreement with the National Science Foundation. 
The W.M. Keck Observatory is operated as a scientific partnership among the California Institute of Technology, the University of California, and the National Aeronautics and Space Administration. The Observatory was made possible by the generous financial support of the W.M. Keck Foundation.
The WIYN Observatory is a joint facility of the University of Wisconsin-Madison, Indiana University, Yale University, and the National Optical Astronomy Observatories. 
The South African Large Telescope (SALT) of the South African Astronomical Observatory is operated by a partnership between the National Research Foundation of South Africa, Nicolaus Copernicus Astronomical Center of the Polish Academy of Sciences, the Hobby-Eberly Telescope Board, Rutgers University, Georg-August-Universit\"at G\"ottingen, University of Wisconsin-Madison, University of Canterbury, University of North Carolina-Chapel Hill, Dartmough College, Carnegie Mellon University, and the United Kingdom SALT consortium.

\bibliography{distpaper}
\bibliographystyle{distpaper}
\addcontentsline{toc}{chapter}{Bibliography}

\end{document}